\documentclass{article}
\usepackage{subfig}
\usepackage{tikz}
\usepackage[]{graphicx}
\usepackage{geometry}
\usepackage{amsmath}
\usepackage{amssymb}
\usepackage{enumitem}
\usepackage[T1]{fontenc}
\usepackage{array}
\usepackage{booktabs}
\usepackage{tablefootnote}
\usepackage{multirow}
\usepackage{colortbl}
\usepackage{float}
\usepackage{fix-cm} 
\usepackage{bm}
\usepackage{siunitx}
\usepackage{gensymb}
\usepackage{hyperref}
\usepackage{parskip}
\usepackage{titling}

\newcolumntype{P}[1]{>{\raggedright\arraybackslash}p{#1}}
\usepackage[nodayofweek]{datetime}
\usepackage[backend=biber, style=nature]{biblatex}
\definecolor{SpringGreen}{rgb}{0.58, 0.77, 0.45}

\title{African swine fever in wild boar: investigating model assumptions and structure}

\author{
Callum Shaw$^{1}$, Angus McLure$^{1}$ and Kathryn Glass$^{1}$}
\date{$^{1}$ National Centre for Epidemiology and Population Health, Australian National University, Canberra, ACT, Australia}
\begin{document}

\maketitle

\begin{abstract}
	African swine fever (ASF) is a highly virulent viral disease that affects both domestic pigs and wild boar. Current ASF transmission in Europe is in part driven by wild boar populations, which act as a disease reservoir. Wild boar are abundant throughout Europe and are highly social animals with complex social organisation. Despite the known importance of wild boar in ASF spread and persistence, there remain knowledge gaps surrounding wild boar transmission. To investigate the influence of density-contact functions and wild boar social structure on disease dynamics, we developed a wild boar modelling framework. The framework included an ordinary differential equation model, a homogeneous stochastic model, and various network-based stochastic models that explicitly included wild boar social grouping. We found that power law functions (transmission $\propto$ density$^{0.5}$) and frequency-based density-contact functions were best able to reproduce recent Baltic outbreaks; however, power law function models predicted considerable carcass transmission, while frequency-based models had negligible carcass transmission. Furthermore, increased model heterogeneity caused a decrease in the relative importance of carcass-based transmission. The different dominant transmission pathways predicted by each model type affected the efficacy of potential interventions, which highlights the importance of evaluating model type and structure when modelling systems with uncertainties. 
\end{abstract}

\section{Introduction}
African swine fever (ASF) is a viral haemorrhagic disease caused by the African swine fever virus (ASFV), which affects both domestic pigs and wild boar populations. There is currently no vaccine, and highly virulent strains of ASFV can result in close to 100\% mortality~\parencite{gallardo2017experimental}, while infection from moderately virulent strains has a lower mortality, ranging from 30 to 70\%~\parencite{sanchez2015update}. ASF outbreaks have the potential to devastate pig production. ASF was first detected in China in August 2018. By mid-2019, ASFV infection had killed 13,355 pigs and a further 1.2 million pigs were culled to prevent continued spread. The estimated economic impact of the outbreak was equal to 0.78\% of China`s 2019 gross domestic product~\parencite{you2021african}. Similarly, outbreaks in wild boar have caused significant population decline~\parencite{european2017epidemiological}. Due to this high potential for serious economic losses, ASF has been declared a notifiable disease by the World Organization for Animal Health~\parencite{oie2010terrestrial}. 

ASF was first identified in Kenya in 1921, after the introduction of European domestic pigs~\parencite{montgomery1921form}. Genotype I ASFV was first reported in Europe in 1957, when it was discovered in Portugal. ASF was subsequently reported in Brazil and a number of Caribbean and European countries~\parencite{costard2009african}. In the summer of 2007, there was an ASF outbreak in Georgia, caused by the highly virulent genotype II ASFV. This genotype quickly spread through the Caucasus region and entered the Russian Federation by November 2007~\parencite{gogin2013african}. Despite control efforts, the virus has since reached Eastern Europe and the European Union, where wild boar have been a key driver in the spread~\parencite{pejsak2014epidemiology, chenais2019epidemiological}. In mid-2018, genotype II ASFV reached north-eastern China~\parencite{zhou2018emergence} and by early 2019, ASF had spread to 25 provinces in China~\parencite{zhou2019african}. Since its detection in China, there have been outbreaks in much of East and South-East Asia\parencite{penrith2020current}, with detection of ASF in Indonesia, Timor-Leste, and Papua New Guinea.

The epidemiology of ASF is complex, with four primary transmission cycles identified~\parencite{chenais2018identification}. The first is the sylvatic cycle that occurs in sub-Saharan Africa, in which ASFV circulates between warthogs and several soft tick species in the \textit{Ornithodoros} genus, without causing disease in the warthog. As there is no horizontal transmission between warthogs, soft ticks are essential for transmission~\parencite{peirce1974distribution}. In the second cycle, ASFV is transmitted between soft ticks and domestic pigs. This has been observed in both the Iberian peninsula~\parencite{perez1994relationship, boinas2011persistence} and sub-Saharan Africa~\parencite{costard2013epidemiology}. The third cycle, the domestic cycle, involves transmission through direct or indirect contact between domestic pigs. Indirect transmission via contamination is possible, as ASFV can remain viable in excretions of infected pigs for days~\parencite{davies2017survival} and viable for months in blood~\parencite{plowright1967stability}. Furthermore, infectious ASFV was found to last several days in various soil matrices~\parencite{carlson2020stability} and in shipped feed~\parencite{stoian2019half}. The final cycle of infection is centred on wild boar and their habitat~\parencite{chenais2018identification}. Wild boar have been an important driver of the recent ASF outbreak in eastern Europe~\parencite{sauter2021african} and South Korea~\parencite{jo2020african}. Environmental transmission is a key component of the wild boar transmission cycle; viable ASFV has been detected in wild boar carcasses for weeks after death~\parencite{fischer2020stability} and wild boar have been observed interacting with conspecific carcasses and the environment around said carcasses~\parencite{probst2017behaviour}. The importance of infected carcasses in the transmission cycle is supported by recent modelling work~\parencite{lange2017elucidating,o2020modelling, pepin2020ecological, gervasi2021african}.

Wild boar are highly social animals and are often observed in groups. Most often, groups are made up of one or more sows and their piglets, but can consist of young males or females~\parencite{choq1996pests}. Conversely, adult male boar are often solitary. This wild boar social structure has been found to influence ASF transmission dynamics~\parencite{pepin2021social}. However, wild boar population ecology is not homogeneous as group size, home range area, density, and growth rates vary between regions.

Previous ASF modelling studies focusing on wild boars have been primarily individual-based (agent-based) models~\parencite{gervasi2021african, pepin2020ecological, halasa2019simulation, lange2017elucidating, ko2023simulating} or a homogeneous ordinary differential equation (ODE) model~\parencite{o2020modelling}. Individual-based models, while often the most accurate abstraction of a system, are computationally expensive and require a significant number of parameters, which may challenging to measure due to the knowledge gaps around ASF transmission in wild boar~\parencite{hayes2021mechanistic}. Conversely, simple ODE models may not properly capture the complex social structure of wild boar populations. In this study, we develop a lightweight, highly adaptable ASF modelling framework that can model ASF dynamics in wild boar in a variety of settings. The framework ranges from simple ODE models to stochastic models that explicitly model wild boar group structure. The objective of the study is to highlight the impact of model assumptions, wild boar population characteristics, and ASF transmission features that impact long-term modelled ASF behaviour. 

\section{Model overview}

\begin{figure}[h]
	\centering
	\resizebox{!}{3.75cm}{
	\begin{tikzpicture}[]
		\small
		\draw [rounded corners=1,fill=SpringGreen!60] (0,1.5) rectangle (1.5,2.5) node[pos=.5] { \Large $S$};
		\draw [rounded corners=1,fill=orange!20] (4,1.5) rectangle (5.5,2.5) node[pos=.5] { \Large $E$};
		\draw [rounded corners=1,fill=red!60] (7,1.5) rectangle (8.5,2.5) node[pos=.5] { \Large $I$};
		\draw [rounded corners=1,fill=SpringGreen!20](11,3) rectangle (12.5,4) node[pos=.5] { \Large $R$};
		\draw [rounded corners=1,fill=red!20] (11,0) rectangle (12.5,1) node[pos=.5] { \Large $C$};
		
		\draw [->] (1.5,2) -- node[above=1mm] {$\beta(I+\omega C)\frac{S}{N + C}$}  ++ (2.4,0);
		\draw [->] (5.5,2) -- node[above=1mm] {$\zeta E$}  ++ (1.4,0);
		
		\draw[-] (7.75,2.5) -- (7.75,3.5);
		\draw[->] (7.75,3.5) -- node[above=1mm] {$(1-\nu) \gamma I$} ++ (3.15,0);
		
		\draw[-] (7.75,1.5) -- (7.75,0.5);
		\draw[->] (7.75,0.5) -- node[above=1mm] {$\nu \gamma I$ + $f_d(N) I$} ++ (3.15,0);

		\draw [->] (-1.5,2) -- node[above=1mm] {$f_b(t,N) N$}  ++ (1.4,0); 
		
		\draw [->] (0.75,1.5) -- node[left = 0mm] {$f_d(N) S$} ++ (0,-1);
		\draw [->] (4.75,1.5) --  node[left = 0mm] {$f_d(N) E$} ++ (0,-1);
		
		\draw [->] (12.5,0.5) --  node[above = 1mm] {$\lambda(t) C$} ++ (1.4,0);
		\draw [->] (12.5,3.5) --  node[above = 1mm] {$f_d(N) R$} ++ (1.4,0);
		
		\draw [->] (11.75,4) -- (11.75,4.5) -- node[below = 1mm] {$\kappa R$}++ (-11,0) -- (0.75,2.6);
	\end{tikzpicture}
}
	\caption{Underlying model structure, used in all tested models. There are five base classes: $S$- susceptible, $E$- exposed, $I$- infectious, $R$- recovered, and $C$- carcasses. In this model, $N$ is the the total living population, $f_b(t,N)$ is the time-dependent birth rate, $f_d(N)$ is the death rate, $\lambda(t)$ is the carcass decay rate, $\beta$ is the transmission rate, $\omega$ is the relative infectiousness of carcasses, $\zeta$ is the latent rate, $\gamma \nu$ is the ASF induced mortality rate, $\gamma (1-\nu)$ is the recovery rate, and $\kappa$ is the waning immunity rate.}
	\label{fig:model_strcutre}

\end{figure}
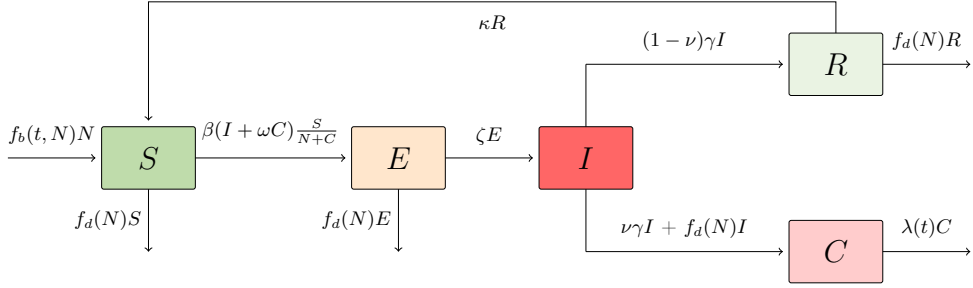

To accurately model an ASFV outbreak in naive wild boar populations, we extended the traditional compartmental susceptible $(S)$, infected $(I)$, recovered $(R)$ model~\parencite{kermack1927contribution, anderson1992infectious}. We included an exposed, non-infectious, class $(E)$ and due to the importance of potential carcass transmission we included a class for the decaying carcasses of pigs that have recently died from ASF $(C)$. For simplicity we did not model piglets; instead we only modelled yearlings and adults pigs, which we combined to form a single adult class. The structure of the model is shown in Fig. \ref{fig:model_strcutre},  where $N$ is the the total living population. $f_b(t,N)$ is the time-dependent birth rate, $f_d(N)$ is the death rate, and $\lambda(t)$ is the carcass decay rate. $\beta$ is the transmission rate, $\omega$ is the relative infectiousness of carcasses compared to live pigs, $\zeta$ is the latent rate, $\gamma \nu$ is the ASF induced mortality rate, $\gamma (1-\nu)$ is the recovery rate, and $\kappa$ is the waning immunity rate.

\subsection{Births}
Seasonality can have a large impact on disease dynamics in ecological systems~\parencite{altizer2006seasonality}. The seasonal cycle of births provides new susceptible individuals and can affect the critical community size~\parencite{peel2014effect}. Seasonality in wild boar births has been observed in many geographic settings~\parencite{caley1993ecology, morreti1995birth, taylor1998reproduction, albrycht2016demographic}. In our study, the total birth rate $f_b(t,N)$ (Eq. \ref{eq:split}) is primarily composed of a raw birth rate $b(t)$, which is scaled to account for population-level density effects. The raw birth rate is modelled with a periodic Gaussian function similar to that used by Peel \textit{et al.}~\parencite{peel2014effect}.
\begin{equation}
	b(t) = b_0 \exp{\bigg(-s\cos\bigg(\frac{\pi (t - \phi_b)}{365}\bigg)^2\bigg)},
	\label{eq:bp}
\end{equation}
where $s$ alters pulse width, $\phi_b$ alters the phase of the pulse, and $k$ is a scaling factor to ensure the correct total yearly birth rate. As piglets are not modelled, the total raw yearly birth rate is the birth rate of young that survive the first year. Therefore
\begin{equation}
	b_0 = \frac{0.5*l_t*l_s*(1-l_\mu)}{\int_0^{365} \exp{\bigg(-s\cos\bigg(\frac{\pi (t - \phi_b)}{365}\bigg)^2\bigg)}dt}.
\end{equation}
We assume there is an even split between male and female pigs, and the birth rate is the product of the average local number of litters per year for yearlings/adults ($l_t$), the average local litter size ($l_s$), and the local probability that a piglet will not die the first year (1-$l_{\mu}$).

\subsection{Deaths and carrying capacity }
The logistic equation is a staple of ecological modelling, used to implement a population-level carrying capacity $(K)$; however, it is not applicable in all contexts. The logistic equation often takes the following form:
\begin{equation}
	\dfrac{dN}{dt} = r N\bigg(1-\frac{N}{K}\bigg), 
	\label{eq:logistic}
\end{equation}
where $N$ is the population size and $r$ is the net growth rate (birth rate $(b)$ - death rate $(d)$). This formulation contains a density independent term $(rN)$ and a density dependent term $(rN^2/K)$. This allows for a net positive rate when $ N <K$ and a net negative rate when $N >K$, in order for the population to return to $K$. Furthermore, when $N=K$, the two terms are equal and the net rate is 0. 

\begin{figure}[h!]
	\centering
	\includegraphics{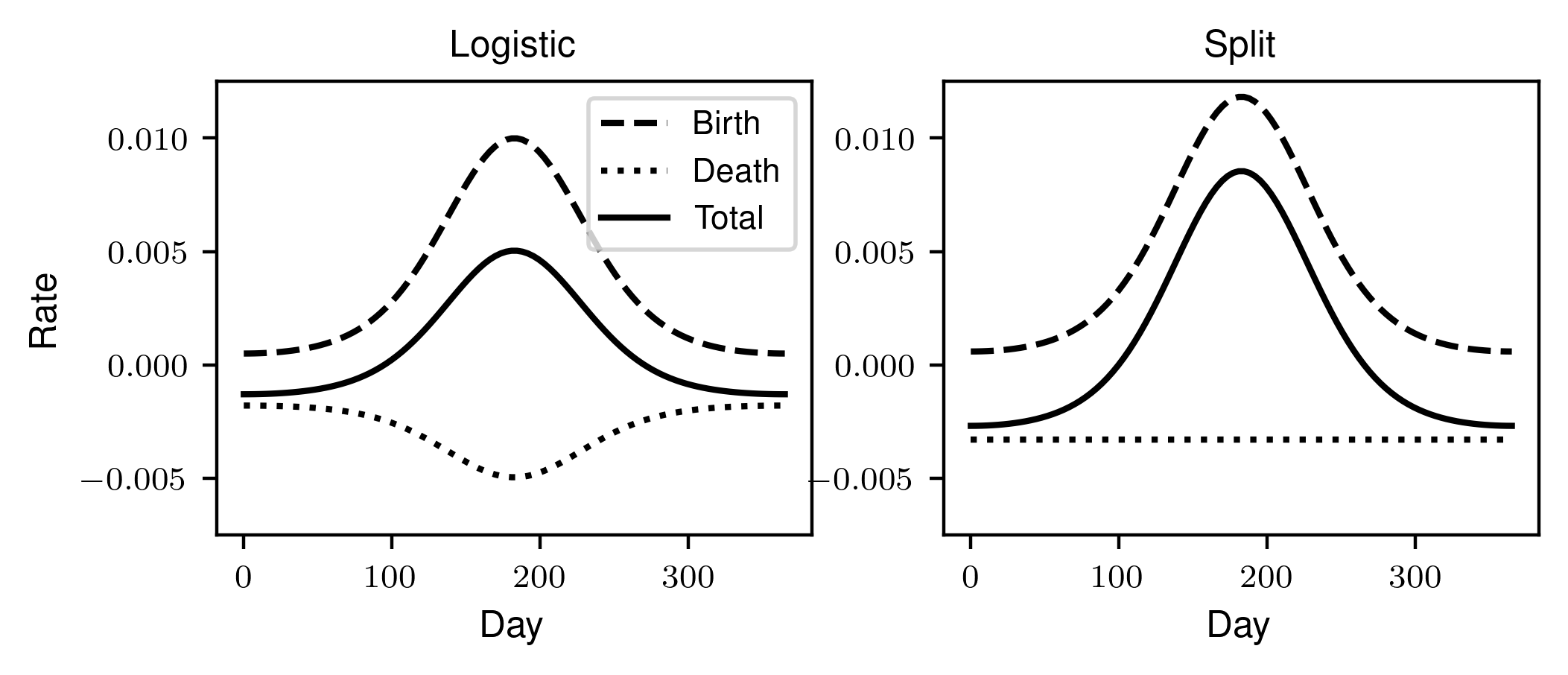}
	\caption{Comparison between the logistic equation and the split equations for the per pig population growth rate. The split model assumes a birth pulse (equation \ref{eq:bp}) with $b_0$ = 0.01, $s$ = 3, and $\phi_b$ = 0, $\mu \approx$ 0.0367, $\sigma$ = 0.75, $\theta$ = 1/2, and $N/K$ = 0.5. The logistic model assumes the same birth pulse, but a death rate ($d$) that is 2/3 the birth rate ($b(t)$).}
	\label{fig:logsplit}
\end{figure}

The first issue is that the basic logistic equation does not allow for seasonality in the population size when $N=K$ as the net rate is 0.  This can be remedied by allowing for seasonal variation in $K$. Furthermore, the logistic equation is a sensible formulation when $r$ is positive; however, unwanted behaviour can occur  if $r$ is negative. If births are time dependent, for example  $b(t)$ has the form shown in equation \ref{eq:bp} and $s>0$, while $d$ is constant and $\min(b(t))< d < \max(b(t))$, $r$ is positive  during the birth pulse peak, yet outside this time $r$ is negative. The second issue with the logistic equation, which occurs both when $K$ is constant or seasonally varying, is that there is an unrealistic timing of deaths. As seen in Fig. \ref{fig:logsplit}, there is a pulse of deaths in the logistic equation that mirrors the pulse in births. In our model, as we did not model piglets (modelled only births that survived to yearling/adult), we would not expect an excess of deaths to occur at the same time as the birth pulse. Furthermore, by concentrating most deaths into a pulse, this formulation may improperly model deaths outside of the birth pulse, as studies have found that adult wild boar deaths are distributed throughout the year~\parencite{caley1993ecology, jezierski1977longevity}.  Therefore, due to the known importance of death and birth rates on disease dynamics~\parencite{altizer2006seasonality}, we formulated another method to implement $K$.

To remedy the issues with equation logistic equation, we assumed that death rate was independent of the birth rate and that both the birth and death rates contain density dependent and independent components. The equations take the form of:

\begin{align}
	\begin{split}
		f_b(t, N) &=  b(t)\left(\sigma+(1-\sigma)\left(\frac{K}{N}\right)^{\theta}\right),\\
		f_d(N)  &=  \mu\left(\sigma+(1-\sigma)\left(\frac{N}{K}\right)^{\theta}\right), 
		\label{eq:split}
	\end{split}
\end{align}
where $b(t)$ is the aforementioned raw birth rate, $\sigma$ is the modifier to determine the ratio of density-independent births/deaths to total births/deaths (1 is purely density independent, whereas 0 is purely density-dependent), $\theta$ is the power of the density-dependent component (assumed to be 1/2), and  $\mu$ is the daily time-independent death rate. As the average simulated model time was comparable to the average wild-boar lifetime, we assumed that $K$ was constant. As such, without ASF, when the population size is equal to $K$, the yearly births equals yearly deaths. If $\theta=1/2$, the birth and death equations in equation \ref{eq:split}, can be combined to form 
\begin{equation}
	\dfrac{dN}{dt} = \sigma r(t)N+(1-\sigma)(b(t)K-\mu N)\bigg(\frac{N}{K}\bigg)^{1/2}.
	\label{eq:splitcomb}
\end{equation}
Equation \ref{eq:splitcomb} allows for seasonal variation in $K$, decoupled births and deaths, and a variable split between the relative strength of density and density-independent processes. The effects of $\sigma$ and its relationship to growth rate are explored in Appendix Section 1.1. As seen in Fig. \ref{fig:logsplit}, with the \textit{split} formulation there is no longer the unwanted pulse of deaths mirroring the birth pulse, instead the deaths are distributed throughout the year. A more complete comparison of the split formulation, logistic equation, and logistic equation with a seasonally varying $K$, for three different ratios of $N/K$, is shown in Appendix Fig. 2.

\subsection{Model structure}
We considered three ASF models based on the model structure shown in Fig. \ref{fig:model_strcutre}. The models were:
\begin{enumerate}
	\item Homogeneous ordinary differential equation model (M1)
	\item Homogeneous stochastic model (M2)
	\item Heterogeneous network-based stochastic model (M3) 
\end{enumerate}

M1 was run with a variable time step and was defined by the following set of equations:
\begin{align}
	\begin{split}
		\dfrac{dS}{dt}&=f_b(t,N) N + \kappa R - f_d(N) S -\beta(I+\omega C)\frac{S}{N + C},\\
		\dfrac{dE}{dt}&=\beta(I+\omega C)\frac{S}{N + C}-(f_d(N)+\zeta)E,\\
		\dfrac{dI}{dt}&=\zeta E - (f_d(N)+\gamma)I,\\
		\dfrac{dR}{dt}&=\gamma(1-\nu)I - (f_d(N)+\kappa)R,\\
		\dfrac{dC}{dt}&=(f_d(N) + \gamma \nu)I -\lambda(t) C,\\
	\end{split}
	\label{eq:ode_sim}
\end{align}
where
\begin{equation}
	\lambda(t) = \lambda_0 + \lambda_1*\cos\left(\frac{2\pi(t+\phi_\lambda)}{365}\right) .
	\label{eq:birthdeath}
\end{equation}

\begin{figure}[h!]
	\centering
	\resizebox{!}{5cm}{
		\begin{tikzpicture}
			
			\draw [fill=SpringGreen!20] (3,3) circle (2.75cm);
			
			\draw [solid,line width = 2pt] (3,3) -- (4.75,3);
			\draw [solid,line width = 2pt] (3,3) -- (3.875,4.75);
			\draw [solid,line width = 2pt] (3,3) -- (3.875,1.25);
			\draw [solid,line width = 2pt] (3,3) -- (2.125,4.75);
			\draw [solid,line width = 2pt] (2.125,4.75) -- (3.875,4.75);
			\draw [solid,line width = 2pt] (2.125,4.75) -- (4.75,3);
			\draw [solid,line width = 2pt] (2.125,1.25) -- (3.875,1.25);
			\draw [solid,line width = 2pt] (2.125,1.25) -- (1.25,3);
			\draw [solid,line width = 2pt] (1.25,3) -- (3.875,1.25);
			\draw [solid,line width = 2pt] (1.25,3) -- (2.125,4.75);
			
			\draw [fill=blue!20](3,3) circle (0.5cm);
			\draw [fill=yellow!20](1.25,3) circle (0.5cm);
			\draw [fill=yellow!20](4.75,3) circle (0.5cm);
			
			\draw [fill=yellow!20](2.125, 1.25) circle (0.5cm);
			\draw [fill=yellow!20](2.125, 4.75) circle (0.5cm);
			
			\draw [fill=yellow!20](3.875, 4.75) circle (0.5cm);
			\draw [fill=yellow!20](3.875, 1.25) circle (0.5cm);
			
		\end{tikzpicture}
	}
	\caption{Network model population structure. Sow groups are yellow, while solitary boar \textit{groups} are blue. Inter-group connections are shown by solid black lines.}
	\label{fig:network}
\end{figure}
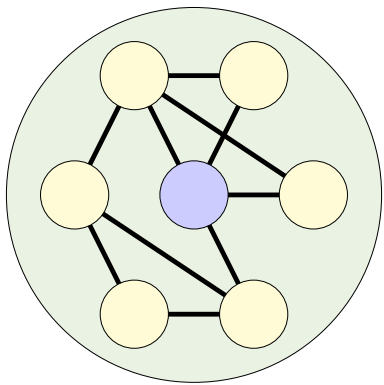

M2 and M3 were run with a Gillespie tau-leaping algorithm~\parencite{cao2006efficient}, which allowed for demographic stochasticity. The models were run with a fixed daily time-step. The underlying structure of the two stochastic models is represented by eleven distinct processes (Appendix Table 1), which included births, deaths, transmission, and disease progression. All models simulated a single population of wild boar. Models M1 and M2 were run on a homogeneous population, while M3 was run on a network containing $N_g$ groups of wild boar. The groups consisted of multiple pigs or a solitary male boar. We assumed that intra-group contact was homogeneous (due to the small average group size), but modelled the complex system of inter-group contacts with a network~\parencite{podgorski2014long}. Nodes were groups and edges represented contact between groups. Therefore, the degree of a group ($k$), was the number of other groups a group could directly interact with. We assumed that, on average, each group had contact with six other groups (${<}k{>}=6 $). Previous wild boar studies found that male boars have larger home-ranges than females~\parencite{saunders1996movements,caley1997movements}. Therefore, we assumed that larger home-ranges of male boar would result in an increased average number contacts.  To include this in the model, the $\chi\%$ of nodes with largest $k$ were assigned to be a solitary male boar, while the remaining nodes were sow groups (of size $N_s$). An indicative population structure of M3 is given in Fig. \ref{fig:network}. 

Network-type is known to influence epidemic dynamics~\parencite{keeling2005networks}; however, as the true network is unknown, we used several algorithms to generate different idealised networks with varying characteristics to investigate the effect of network type. The following three networks were tested:
\begin{enumerate}
	\item Erd{\H{o}}s–R{\'e}nyi random network (RN)~\parencite{erdHos1960evolution}
	\item Barab{\'a}si–Albert scale-free network (SF)~\parencite{albert2002statistical}
	\item Watts–Strogatz small-world random network (SW)~\parencite{watts1998collective}
\end{enumerate}

The chosen networks cover a range of attributes. In RN networks, the degree distribution of each group, provided the number of groups modelled is sufficiently large, follows a Poisson distribution and clustering is low.  In SF networks, the degree distribution follows a power-law ($P(k) \propto k^{-3}$), such that ${<}k{>}$ is finite, but the variance is infinite. Therefore, most groups have low connectivity, but some groups (solitary boars) are highly connected and act as key hubs (nodes with high betweenness centrality) in the network. The average clustering in the SF networks generated for this study is low. The SW network in this study was run with a rewiring probability of 0.2, which allowed the network to have high clustering. 

\subsection{Transmission coefficient}
There are two transmission pathways for ASF included in the model: infection from live pigs, or infection from infectious carcasses of pigs that have recently died from ASF. In the homogeneous models, M1 and M2, both pathways are governed by a single transmission rate, $\beta_h$. In M3, infection can come from within a pig's group or from a connected group, hence there are two rates -- intra-group transmission ($\beta_i$) and inter-group transmission ($\beta_o$). To determine the role of carcasses in transmission, infection from carcasses is set to be proportional to $\omega \beta_x$, where $\omega$ is the relative infectiousness of carcasses to live pigs, a fitted parameter. 

\begin{figure}[htb]
	\centering
	\includegraphics{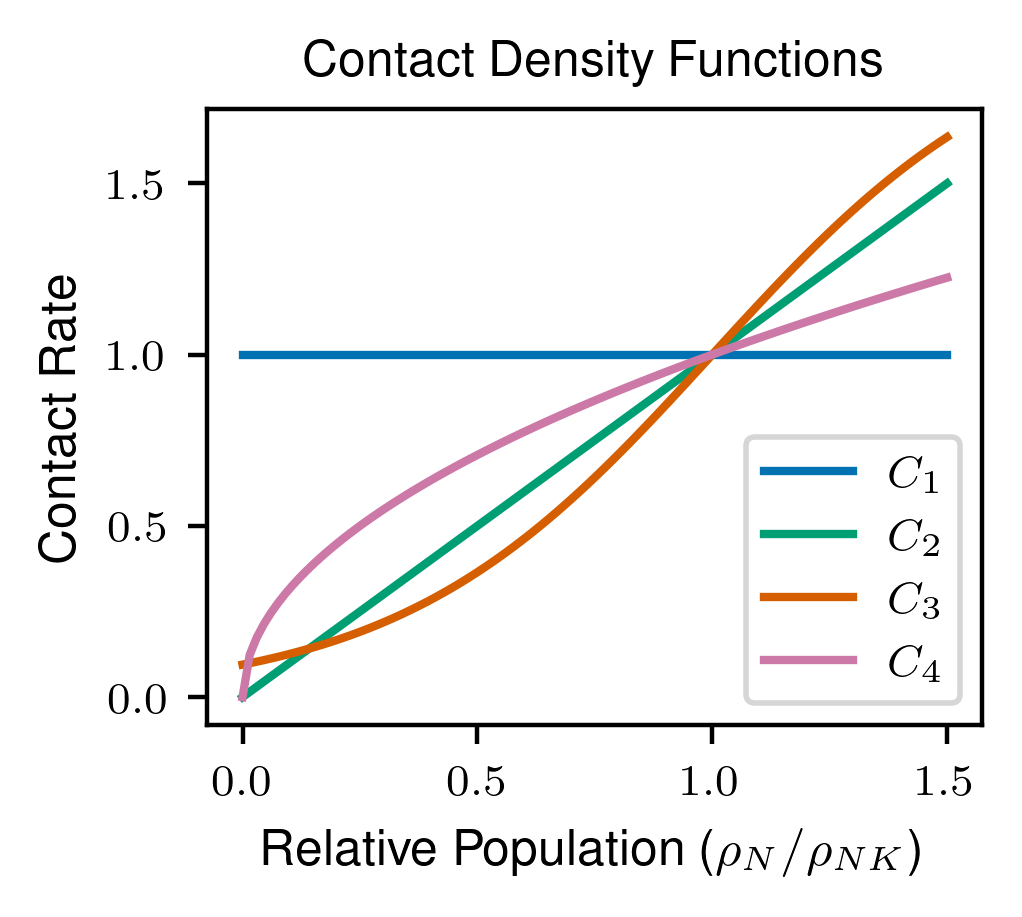}
	
	\caption{Plots of the four different contact-density functions tested in the models. The frequency-based contact function ($C_1$) is in blue, the density-based contact function ($C_2$) is in green, the sigmoid contact function ($C_3$) is in orange, and the power-law contact function ($C_4$) is in pink.}
	
	\label{fig:plotdiff}
\end{figure}

In wild boar, higher densities likely lead to high contact rates~\parencite{cowled2008review}, however, the exact contact density function is not known. A positive relationship between density and disease spread has been noted at the population level, in both classical swine fever studies~\parencite{artois2002classical,rossi2005incidence,kramer2009individual} and ASF studies~\parencite{nurmoja2017development, podgorski2020spatial, o2020modelling}. Furthermore, power functions have been shown to offer a realistic scaling of transmission rate with density in ecological settings~\parencite{borremans2017shape, smith2009host, hu2013scaling}. Therefore, similar to Borremans \textit{et al.}~\parencite{borremans2017shape}, we tested a number of contact-density functions that scaled with density. We set $\beta = C(\rho_N)p$, where $C(\rho_N)$ is the contact rate, a dimensionless scaling factor that is a function of population level density ($\rho_N$), and $p$ is the transmission rate. We tested four formulations of  $C(\rho_N)$ for $\beta_h$ and $\beta_o$; these relationships are given in Table \ref{table:transmission_functions} and plots of how each formulation scales with density are shown in Fig. \ref{fig:plotdiff}.
\begin{table}[htb]
	\centering
	\begin{tabular}{lll} 
		\toprule
		\bf{Function \#} & \bf{Transmission-type} & \bf{Description}   \\ \midrule
		
		$C_1$ & Frequency-based & $C(\rho_N) = 1$ \\
		$C_2$ & Density-based & $C(\rho_N) =  \rho_N/\rho_{NK}$ \\
		$C_3$ & Sigmoid & $C(\rho_N) =  \tanh(1.5 \rho_N/\rho_{NK}-1.5) + 1$\\
		$C_4$ & Power-law & $C(\rho_N) = (\rho_N/\rho_{NK})^{1/2}$ \\
		\bottomrule
		
	\end{tabular}
	\caption{The four different contact-density functions tested in the models. Here, $C(\rho_N)$ is the contact-density function, $\rho_N$ is the wild boar density (total pigs per unit area), and $\rho_{NK}$ is the wild boar density at $K$.}
	\label{table:transmission_functions}
\end{table} 
While the model assumes that inter-group transmission is dependent on density, intra-group transmission is density independent. Bouma \textit{et al.}~\parencite{bouma1995transmission} found that intra-group pseudo-rabies transmission rates, on the group scale, were independent of population density. Therefore, the density-contact function for intra-group transmission was assumed to be frequency-based. 
\section{Fitting models to recent outbreak data}
\subsection{Parameters and model fitting}
All models relied on two sets of parameters: transmission parameters which were assumed the same across populations $\{ p_h, p_o, p_i, \omega,$$\sigma,$$\theta,$$\zeta,$$\gamma,$$\kappa,$$\nu \}$, and population and location specific parameters that likely varied by region $\{K, \rho_{NK},$$ {<}k{>}, $ $N_g, N_s, $ $ \chi, $ $ N_{ly}, $ $ N_{ls}, $ $ \mu_{py}, $ $ \mu,  $ $\phi_b, $ $ \lambda_0, $ $ \lambda_1,  $ $\phi_\lambda\}$. Transmission parameters were estimated by fitting our candidate models to recent outbreak data from the Baltic states and Poland. Transmission parameters are given in Table \ref{table:locin_params}, while population parameters are given in Table \ref{table:baltic_params}.

\begin{table}[htb]
	\centering
	\begin{tabular}{ m{2cm}  m{4.5cm}   m{2cm}  m{3.5cm}} \toprule
		\bf{Parameter} & \bf{Description} & \bf{Value} & \bf{Source}  \\ \midrule
		$p_h$ & Homogeneous transmission rate (M1 \& M2)  & Fitted & -\\
		$p_o$ & Inter-group transmission rate (M3) &  Fitted & - \\
		$p_i$ & Intra-group transmission rate (M3) & Fitted & -\\	
		$\omega$ & Relative infectiousness  of carcasses to live pigs & Fitted & -\\
		$\sigma$ & Proportion of density independent births \& deaths to total births \& deaths & 0.75 &~\parencite{massei2015wild,pascual2022wild} \\
		$\theta$ & Power coefficient of density-dependent births \& deaths & 0.5 & Assumption \\
		$1  / \zeta$ & Latent period & 6 days &~\parencite{guinat2014dynamics, guinat2018inferring, ewing2022exact} \\
		$1 / \gamma$ & Infectious period & 8 days &~\parencite{de2013transmission,guinat2018inferring,ewing2022exact} \\ 
		$1 / \kappa $ & Waning immunity period & 180 days$^1$ \\
		$\nu$ & Lethality & 0.95 &~\parencite{gallardo2017experimental, european2021asf} \\
		\bottomrule
	\end{tabular}
	\caption{Transmission based model parameters. $^1$Immunity duration is a current knowledge gap and requires further investigation~\parencite{blome2020african}. Sereda \textit{et al.}~\parencite{sereda2020protective} found that immunisation with attenuated ASFV strains offered at least four months protection from virulent ASFV strains.}
	\label{table:locin_params}
\end{table} 

\begin{table}[htb]
	\centering
	\begin{tabular}{ m{2cm}  m{4.5cm}   p{2cm}  m{3.5cm}}\toprule
		
		Parameter & Description & Value & Source  \\\midrule
		$K$ & Carrying capacity of modelled region & 5000 pigs & Assumption \\
		$\rho_{NK}$ & Initial wild boar density & 2.8 pigs km$^{-2}$&~\parencite{pascual2022wild} \\
		${<}k{>}^2$ & Average number of inter-group connections & 6 connections & Assumption  \\
		$N_g^2$ & Number of groups  & 1000 groups & Assumption \\
		$N_s^2$& Sow group size & 6 pigs& \parencite{jkedrzejewski1992wolf, poteaux2009socio, pascual2022wild} \\
		$\chi^2$ & Ratio of solitary boar groups to total groups & 0.2 &~\parencite{jkedrzejewski1992wolf}\\
		$N_{ly}$ & Number of yearly litters & 0.9 litters& \parencite{mauget1982seasonality,bieber2005population}\\
		$N_{ls}$ & Litter size & 5.8 pigs&  \parencite{albrycht2016demographic, bieber2005population, frauendorf2016influence, pascual2022wild}\\
		$\mu_{py}$ & First year mortality & 0.5 & \parencite{jezierski1977longevity,bieber2005population, pascual2022wild}\\
		$\mu$ & Death rate at $K$ & 0.0036 day$^{-1}$ & $0.5N_{ly}N_{ls}(1-\mu_{py}) /365 $\\
		$s$ & Birth pulse width & 3 & \parencite{albrycht2016demographic}\\
		$\phi_b$ & Birth pulse offset & 75 days& \parencite{albrycht2016demographic}\\
		$1 / \lambda_0 $ & Base carcass decay period & 60 days& Local weather \& \parencite{fischer2020stability} \\
		$1 / \lambda_1 $ & Seasonal change in carcass decay period & 30  days& \parencite{fischer2020stability}  adjusted for local conditions \\
		$\phi_\lambda$ & carcass decay offset & 0 days& Local weather\\
		\bottomrule
		
	\end{tabular}
	\caption{Population model parameters selected to reflect the Baltic region. $^2$ Parameters only used in heterogeneous network model.}
	\label{table:baltic_params}
\end{table} 

The three models were fitted using an approximate Bayesian computation with a sequential Monte Carlo algorithm (ABC-SMC). Each model had an initial population of 5000 wild boar, which was the assumed carrying capacity of the modelled region. All models assumed that ASFV was introduced during the summer, as wild boar tend to have larger home ranges during this season~\parencite{keuling2008annual}. In M1 and M2, ASF was initialised with 1\% prevalence, while in M3 ASF was seeded in five connected groups. This high number was chosen to minimise the occurrence of minor outbreaks in the stochastic models~\parencite{whittle1955outcome}. In M1 and M2, which assumed homogeneous populations, the fitted parameters were $p_h$ and $\omega$. In M3 the fitted parameters were $p_i$, $p_o$, and $\omega$. All parameters had semi-informative uniform priors and fitting simulations were run for six years. Model fit was judged against summary statistics. The following summary statistics were chosen from the Baltic and Polish ASF outbreaks:
\begin{itemize}
	\item Prevalence of ASFV several years post introduction is 1.5\% (95\% CI [1\%, 2\%])~\parencite{schulz2022eight, pautienius2020african}. 
	\item  75\% decline in wild boar density during the epidemic (95\% CI [65\%, 85\%])~\parencite{european2017epidemiological,schulz2019epidemiological,schulz2021african}.
	\item Infection peak occurs 180 days after the first detected case (95\% CI [120 Days, 240 Days]) ~\parencite{european2017epidemiological}. As there was no surveillance in the model, we assumed the first detected case would occur at 5\% prevalence.
\end{itemize}

To assess the effect of contact-density functions, every model was fitted with the four contact functions (Table \ref{table:transmission_functions}). To analyse the effect of network structure, M3 was also fitted with the three network types (RN, SF, and SW).

\subsection{M1 \& M2 Model fitting results}
In both M1 (ODE) and M2 (homogeneous stochastic) models, the posterior distributions of the fitted parameters for each contact-function sub-model show largely the same behaviour. Properties of the posteriors for $p_h$ and $\omega$ are given in Appendix Table 2 and plots of the M1 and M2 posterior distributions are given in Appendix Fig. 6 and 7. To asses the quality of fit of M1 and M2, for each contact-density formulation, we simulated both M1 and M2 using draws from the fitted posterior distributions. Each sub-model was run 10,000 times and the simulations ran for six years. We then computed summary statistics for each run and checked the goodness of fit between the simulated and observed summary statistics. 

\begin{figure}[htb]
	\centering
	\includegraphics{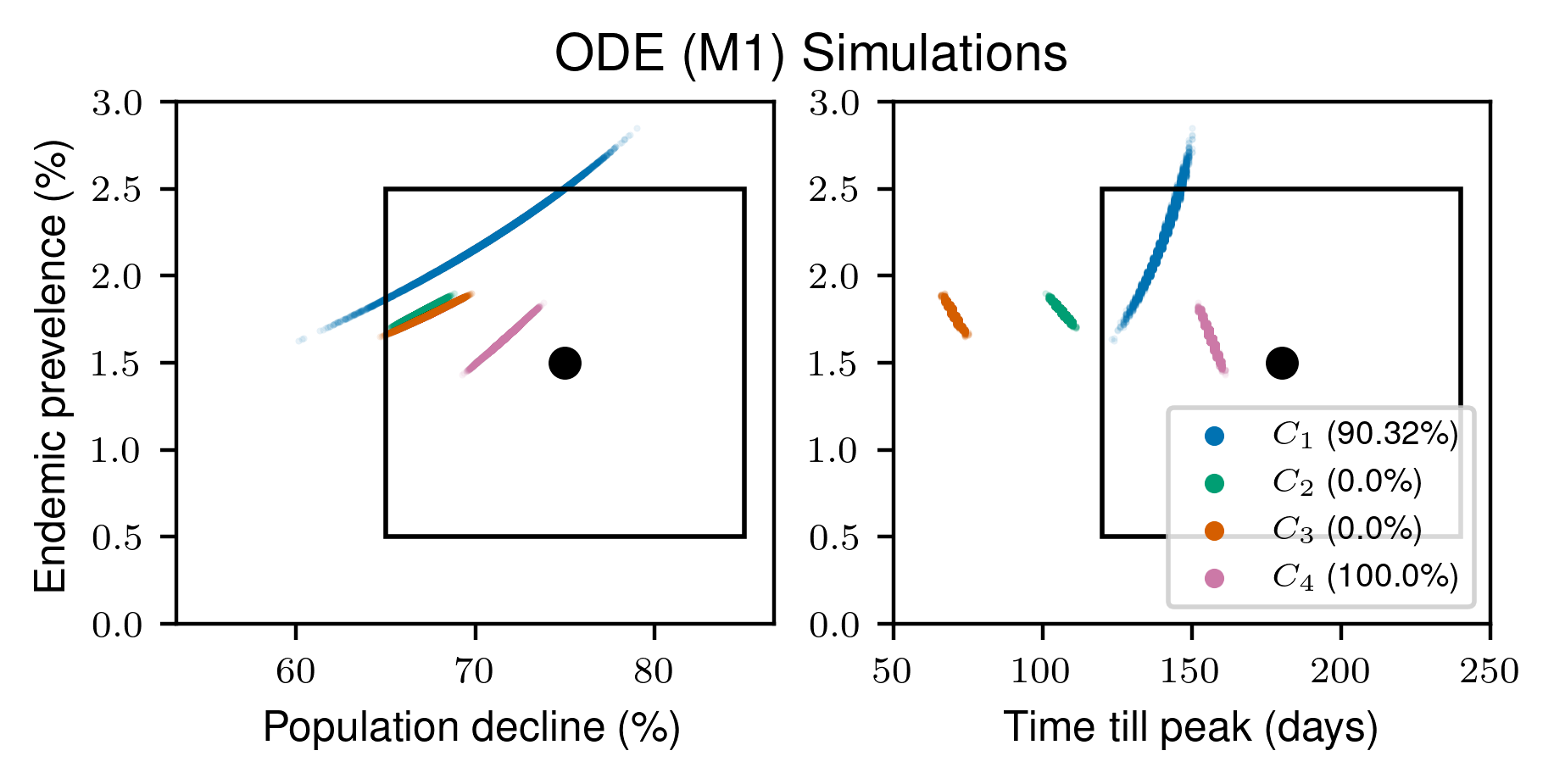}
	\caption{Comparison between the three simulated summary statistics and the observed summary statistics for ODE model (M1). The plot on the left compares statistics S1 and S2, while the right plot compares S1 and S3. The point estimates of the observed summary statistics are represented by the black dot, and the associated 95\% confidence regions by the black rectangles. Results from the frequency-based contact function ($C_1$) are in blue, from the density-based contact function ($C_2$) in green, from the sigmoid contact function ($C_3$) in orange, and from the power-law contact function ($C_4$) in pink. All models were simulated 10,000 times. In brackets in the legend are the proportion of simulations where all three simulated summary statistics were within the 95\% confidence interval of all observed summary statistics.}
	\label{fig:odesims}
\end{figure}

The results for the M1 simulations are given in Fig. \ref{fig:odesims}. The best fitting contact-density function was the power-law relationship $(C_4)$, where 100\% of the simulated summary statistics were within the 95\% confidence intervals of all observed summary statistics. The next most effective contact-density function was $C_1$, where 90\% of simulated summary statistics were within the 95\% confidence interval of all observed summary statistics. No simulated summary statistics for the density-based transmission ($C_2$) and the sigmoid transmission ($C_3$) were within the 95\% confidence interval of all observed summary statistics. For contact functions $C_2$ and $C_3$, the endemic prevalence and population decline were within the 95\% confidence intervals, but ASF spread more quickly than observed in outbreak data with the peak in infections occurring prematurely (on average in 106 days for $C_2$ and in 69.6 days for $C_3$). 

\begin{figure}[htb]
	\centering
	\includegraphics{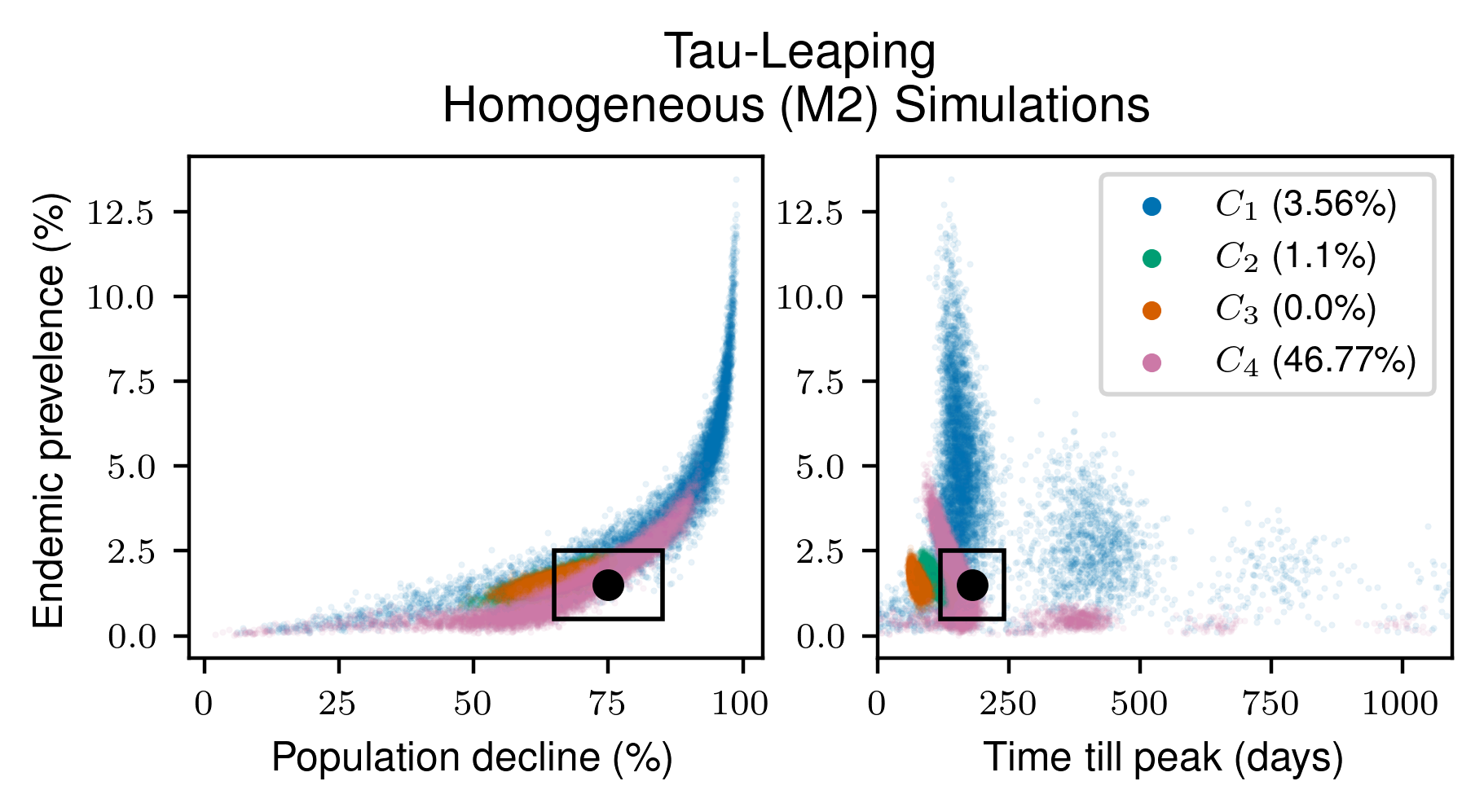}
	\caption{Comparison between the three simulated summary statistics and the observed summary statistics for the stochastic homogeneous model (M2). The plot on the left compares summary statistics S1 and S2, while the right plot compares S1 and S3. The observed summary statistic is represented by the black dot, and its associated 95\% confidence interval by the black rectangle. Results from the frequency-based contact function ($C_1$) are in blue, the density-based contact function ($C_2$) in green, the sigmoid contact function ($C_3$) in orange, and the power-law contact function ($C_4$) in pink. Each contact density formulation model was simulated 10,000 times. In brackets in the legend are the proportion of simulations where all three simulated summary statistics were within the 95\% confidence interval of all observed summary statistics.}
	\label{fig:tauhsims}
\end{figure}

While similar to the M1 results, the stochastic nature of M2 caused a greater spread in simulated summary statistics (Fig. \ref{fig:tauhsims}). Contact function $C_4$ again produced the most accurate model, with approximately 47\% of simulated summary statistics lying within the 95\% confidence intervals of all observed statistics and ASFV was endemic in over 97\% of simulations after six years. All other contact functions produced a poor fit and were largely unable to reproduce the observed summary statistics. Only 3\% of simulated $C_1$ summary statistics were within all 95\% confidence intervals and after six years ASFV was endemic in 70\% of simulations. When endemic, prevalence was often too high; a third of endemic simulations had an average ASF prevalence greater than 5\%.

\begin{figure}[htb]
	\centering
	\includegraphics{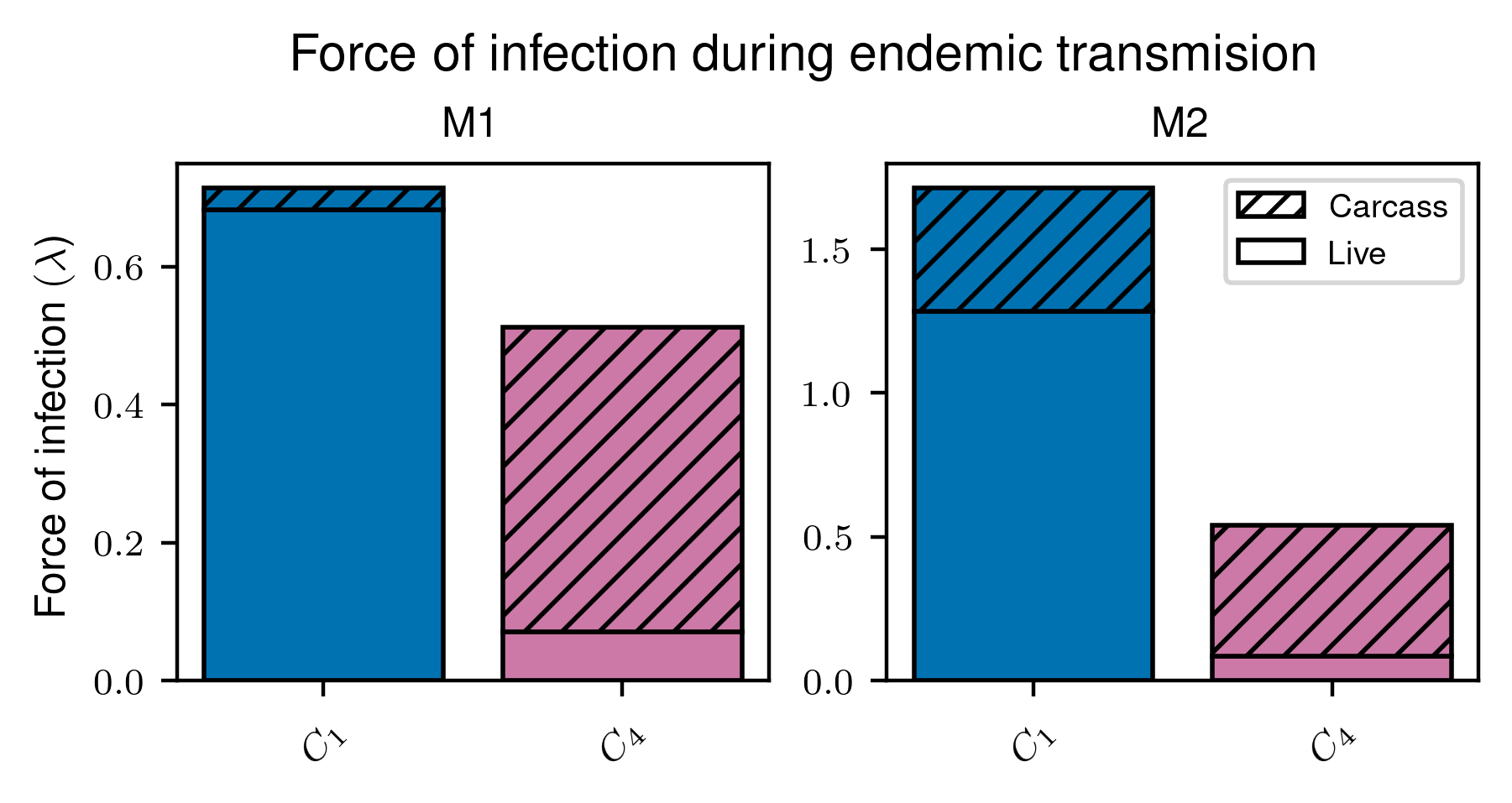}
	\caption{The yearly force of infection ($\bm{\lambda}$) from carcasses and live pigs, computed for the ODE model (M1) and the homogeneous tau-leaping model (M2), with both frequency-dependent transmission (blue) and the power-law contact-density function (pink).  The force of infection was calculated during the endemic phase of the outbreak and results have been filtered to only include endemic simulations.}
	\label{fig:forcem1m2}
\end{figure}

In the homogeneous models (M1), the two most accurate density-contact functions, $C_1$ and $C_4$, predicted different dominant transmission pathways. To analyse the various routes of ASF transmission, the aforementioned simulations were analysed to calculate the force of infection ($\bm{\lambda}$) from both live animals ($\bm{\lambda_l}$) and from carcasses ($\bm{\lambda_c}$). The force of infection was only calculated in endemic simulations during the endemic phase of transmission, between years three and six. Results are given in Fig. \ref{fig:forcem1m2}. In this homogeneous model (M1), the $C_1$ sub-model predicted $\bm{\lambda_l} \approx 20 \bm{\lambda_c}$, while the converse was true in the $C_4$ sub-model, where $\bm{\lambda_l} \approx \frac{1}{6} \bm{\lambda_c}$. Therefore, transmission in the $C_4$ sub-model was primarily driven by carcasses, while in the $C_1$ sub-model transmission was dominated by live pigs and there was comparatively little carcass-based transmission. Furthermore, the total force of infection in the $C_1$ sub-model was 36\% larger than in the $C_4$ sub-model.

In the stochastic model (M2), there was a similar, albeit lesser, dichotomy in transmission predicted by $C_1$ and $C_4$. The $C_1$ sub-model predicted $\bm{\lambda_l} \approx 3 \bm{\lambda_c}$, while in the $C_4$ sub-model, $\bm{\lambda_l} \approx \frac{1}{5} \bm{\lambda_c}$. The total force of infection of the $C_1$ sub-model was 220\% larger than in the $C_4$ sub-model and was 140\% greater than in the M1 $C_1$ sub-model. This high force of infection may stem from the inability of the M2 $C_1$ sub-model to accurately reproduce the observations, as a number of simulations achieved too high an endemic prevalence, which would have caused an overestimation of $\bm{\lambda}$. The total force of infection calculated from the most accurate models (M1 $C_1$, M1 $C_4$, and M2 $C_4$ sub-models) was comparable to that calculated by Loi \textit{et al.}~\parencite{loi2020mathematical}.

The fitting procedure outlined above was repeated for a logistic equation based ODE (LM1) and homogeneous tau-leaping model (LM2) to quantify the effect of using the \textit{split} equation over the logistic equation on ASF dynamics. For LM1, the $C_1$ and $C_2$ functions could accurately reproduce the observed summary statistics; however, in LM2 no contact-density sub-model was able to adequately replicate the summary statistics. The results of the logistic-based simulations are given in Appendix Section 1.2.

\subsection{M3 fitting results}
For the heterogeneous tau-leaping model (M3) an ensemble of six sub-models were fitted (contact functions $C_1$ and $C_4$ run on each of the three network types). $C_2$ and $C_3$ sub-models were excluded due to their poor fit in M1 and M2. The properties of the posterior distributions of intra-group transmission ($p_i$), inter-group transmission ($p_o$), and relative infectivity of carcasses ($\omega$) are given in Appendix Table 3, while plots of the posterior distributions are given in Appendix Fig. 8. 

To assess the goodness of fit of the six M3 models, we again compared summary statistics from simulations using their fitted posteriors to the observed summary statistics. Each model was simulated 1,000 times and the results are given in Fig. \ref{fig:tausims} (Full results in Appendix Table 4). Unsurprisingly, the increased heterogeneity in M3 caused greater variability in summary statistics. Across all networks, the $C_4$ sub-models were better able to reproduce the observed summary statics than the $C_1$ sub-models. Furthermore, $C_4$ sub-models could better model endemicity; after six years, ASFV was endemic in 51.5\% of $C_1$ simulations and 62\% of $C_4$ simulations. There were also notable differences between networks. SW networks, for both contact functions, had the lowest proportion of simulated summary statistics within the 95\% confidence intervals of all observed summary statistics. $C_4$ simulations on both SF and RN networks produced similar levels of fit, while a greater proportion of SF $C_1$ simulations were within the observed summary statistic confidence intervals. 

Compared to the homogeneous stochastic model (M2), depending on the contact-function, the M3 model offered comparable or better levels of fit. The M3 $C_1$ sub-models had an up to 7 times greater proportion of simulations within all observed confidence intervals than the poorly fitting M2 $C_1$ sub-model. In contrast, the M2 $C_4$ sub-model had, on average, a 1.22 times greater proportion of simulations within the observed confidence intervals than the M3 $C_4$ sub-models. For completeness the basic reproduction number ($R_0$) and the effective reproduction number ($R_{eff}$) for M1 and M2 were calculated. Results are presented in Appendix Section 1.6.
\begin{figure}[htb]
	\centering
	\includegraphics{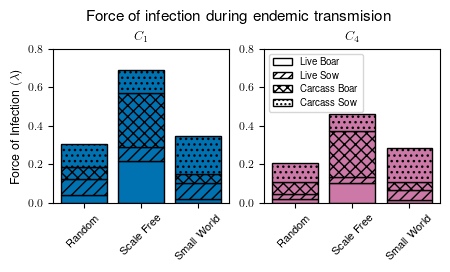}
	\caption{The force of infection ($\bm{\lambda}$) calculated for the tau-leaping heterogeneous model (M3).  On the left, in blue, are sub-models that assume frequency-based density-contact function ($C_1$) and the right plot, in pink, are sub-models that assume a power-law density-contact function ($C_4$). The unhatched bars are force of infection from live boars, the diagonal hatch bars are the force of infection from live sow, the cross-hatch bars are force of infection from boar carcasses, and the dot-hatch bars are force of infection from sow carcasses. The force of infection was calculated during the endemic phase of the outbreak and results have been filtered to only include endemic simulations.}
	\label{fig:forcem3}
\end{figure}
\begin{figure}[htp]
	\centering
	\includegraphics{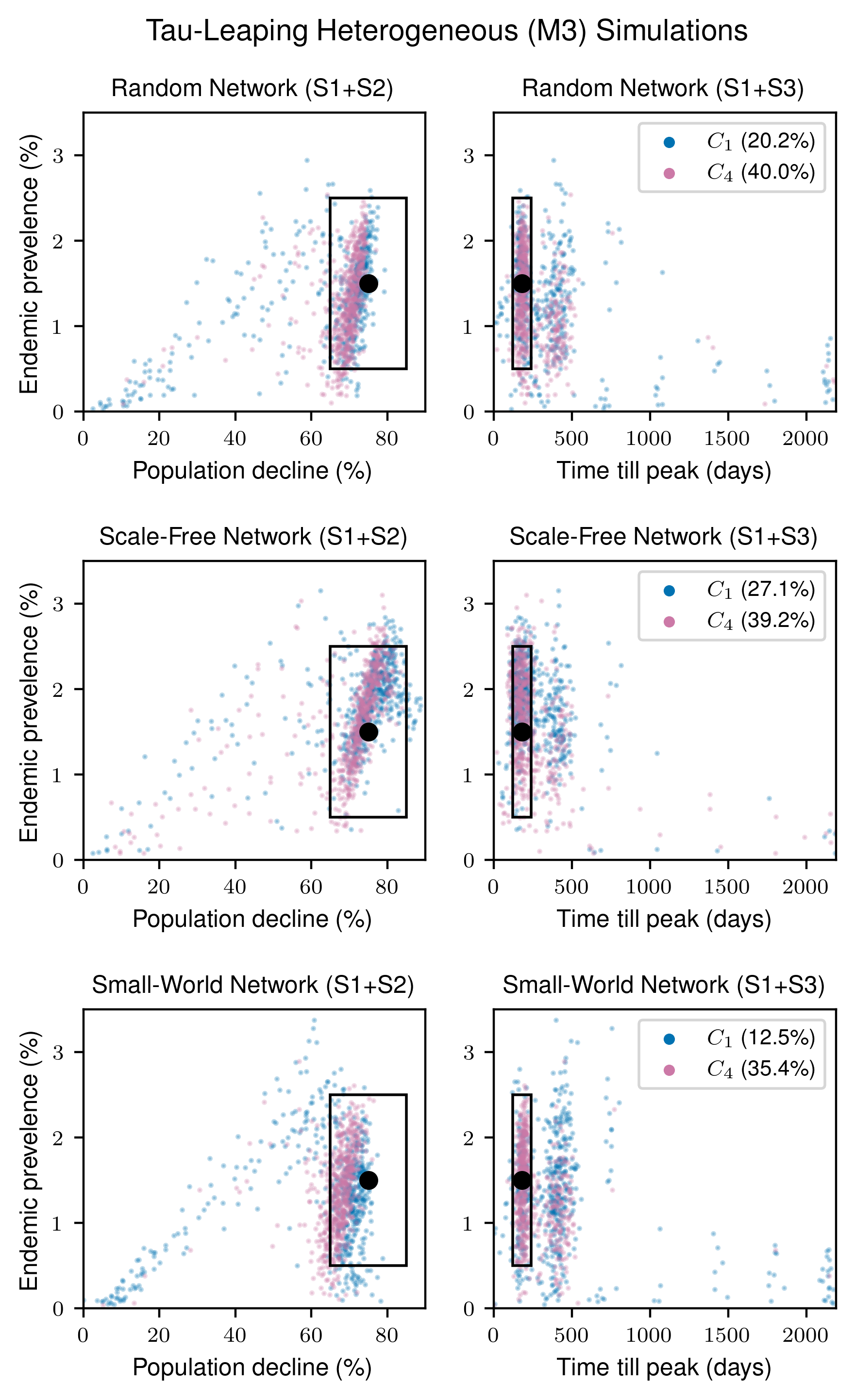}
	\caption{Comparison between the three simulated summary statistics and the observed summary statistics for the tau-leaping heterogeneous model (M3). Each row is the M3 model run on a different network. The left plots compare summary statistics S1 and S2, while the right plots compare S1 and S3. The observed summary statistic is represented by the black dot, and its associated 95\% confidence interval by the black rectangle. Results from the frequency-based contact function ($C_1$) are in blue and from the power-law contact function ($C_4$) are in pink. Due to computational limitations, each model was simulated 500 times. In brackets in the legend are the proportion of simulations where all three simulated summary statistics lie within the 95\% confidence interval of all observed summary statistics.}
	\label{fig:tausims}
\end{figure}

To analyse the different routes of ASF transmission in M3, the six sub-models were again run from their fitted posteriors, simulated 1,000 times, and the force of infection was calculated during the endemic phase of the outbreak (years 3-6). We were able to decompose the force of infection into components from living pigs and pig carcasses, and from solitary boar $(\bm{\lambda_{b}})$ or sow groups $(\bm{\lambda_{s}})$. The results for inter-group transmission are given in Fig. \ref{fig:forcem3} (Full results in Appendix Table 5). As the primary focus of the heterogeneous model was to investigate how ASF spreads through the network and because the force of infection has limited utility within groups (due to small sow group size and solitary boar groups), the force of infection was only computed for inter-group transmission. Consequently, the magnitudes of the calculated forces are not directly comparable to those from the homogeneous models (M1 and M2). 

Infected carcasses were the dominant pathway of transmission for both contact functions ($C_1$ and $C_4$), unlike M1 and M2 where transmission from live pigs dominated in $C_1$ sub-models. However, similar to what was observed in M1 and M2, the total force of infection in the $C_1$ sub-models was equal or greater than that for $C_4$ sub-models. Network structure was also found to influence the force of infection, particularly the relative influence of sows or boar. In the SW models, relative transmission from boars was lowest, with $\bm{\lambda_b} \approx 0.25 \bm{\lambda_s} $. In RN models, while sows were still the primary drivers of infection, boars had a greater influence on ASF spread as $\bm{\lambda_b} \approx 0.56 \bm{\lambda_s} $. Conversely, the SF models were dominated by boar-based transmission as $\bm{\lambda_b} \approx 2.69 \bm{\lambda_s} $. Therefore, in the SF models, although boars only comprised $4\%$ of the population, they were responsible for approximately 70\% of inter-group transmission. This large influence of boars is explained by the SF network structure; the highly connected boars ($k \gg {<}k{>}$) act as hubs in the group-network and drive inter-group transmission. 

\section{Intervention}
\begin{figure}[htpb]
	\centering
	\includegraphics{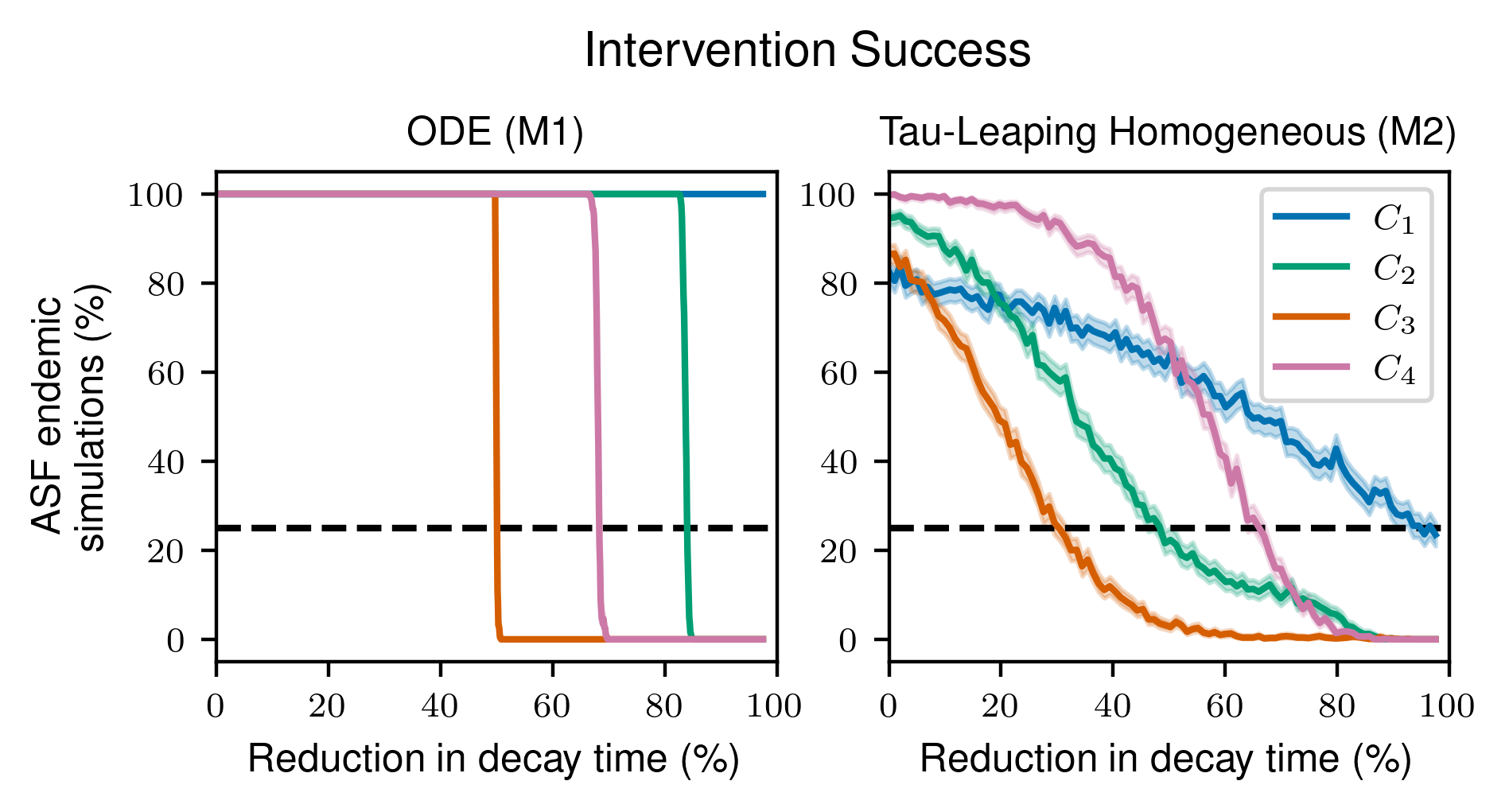}
	\caption{The effectiveness of a carcass removal intervention in both the ODE model (M1) and the tau-leaping homogeneous model (M2). Each model investigates the impact of different assumed contact-density functions: frequency-based ($C_1$) in blue, density-based $(C_2)$ in green, sigmoid-based ($C_3$) in orange, and power-law based ($C_4$) in blue.}
	\label{fig:m1m2_inter}
\end{figure}

To explore the impact of model choice on potential programmatic outcome, we simulated a potential ASF intervention. Specifically, we analysed the efficacy of removing wild boar carcasses to prevent endemic ASFV transmission, as Guinat \textit{et al.}~\parencite{guinat2017effectiveness} and Gervasi \& Guberti~\parencite{gervasi2022combining} found that carcass removal can be a highly effective, albeit potentially impractical, intervention. In our simulations, the intervention started 90 days after ASFV was seeded in the population and continued for three years. The intervention was modelled by reducing the carcass decay period ($1/\lambda$) by multiplicative factor ($\alpha$), where $\alpha \in \left[0,0.975\right]$. While a 97.5\% reduction in decay time is likely unachievable in practice, this upper limit was considered for illustrative purposes. The effect of the intervention was tested for all combinations of model types (M1-M3), contact-density functions ($C_1$-$C_4$), and networks (RN, SF, and SW). For the two homogeneous models, M1 and M2, each magnitude of $\alpha$ tested was simulated 1000 times. The results for M1 and M2 are given in Fig. \ref{fig:m1m2_inter}.

Results from M1 highlight potential limitations with simple homogeneous ODE models. We found that in the $C_1$ sub-model, carcass removal did not reduce the number of simulations with endemic ASF after three years; all simulations had 100\% endemicity after the intervention period. The lack of intervention success occurs because carcass-based transmission in $C_1$ is negligible. The remaining three ODE models exhibited a sudden transition, from ASFV being endemic in 100\% of simulations to ASFV being endemic in no simulations. This bifurcation required an approximate reduction in decay period of 50\% for $C_3$ sub-models, 68\% for $C_4$ sub-models, and 84\% for $C_2$ sub-models. The abrupt shift between the two states is unrealistic and may give a false prediction of potential programmatic success. 

M2, which included demographic stochasticity, did not display the sharp transition between endemicity and extinction that occurred in M1. Instead, compared to M1, all sub-models were characterised by a gradual increase in the number of simulations with ASFV extinction with an increase in the magnitude of $\alpha$. ASFV extinction in 75\% of simulations required an average reduction in the decay period of 97\% for $C_1$ sub-models, 66\% for $C_4$ sub-models, 49\% for $C_2$ sub-models, and 30\% for $C_3$ sub-models. Unlike in M1, the $C_1$ sub-model did respond to the intervention; however, again due to the negligible levels of carcass based transmission, the required intervention was large and would probably not be practically achievable. The addition of stochasticity changed the ordering of the intervention efficacy of sub-models. Importantly, the simple addition of stochasticity removed the sudden bifurcation present in M1 and produced more realistic responses to the intervention.  

\begin{figure}[htpb]
	\centering
	\includegraphics{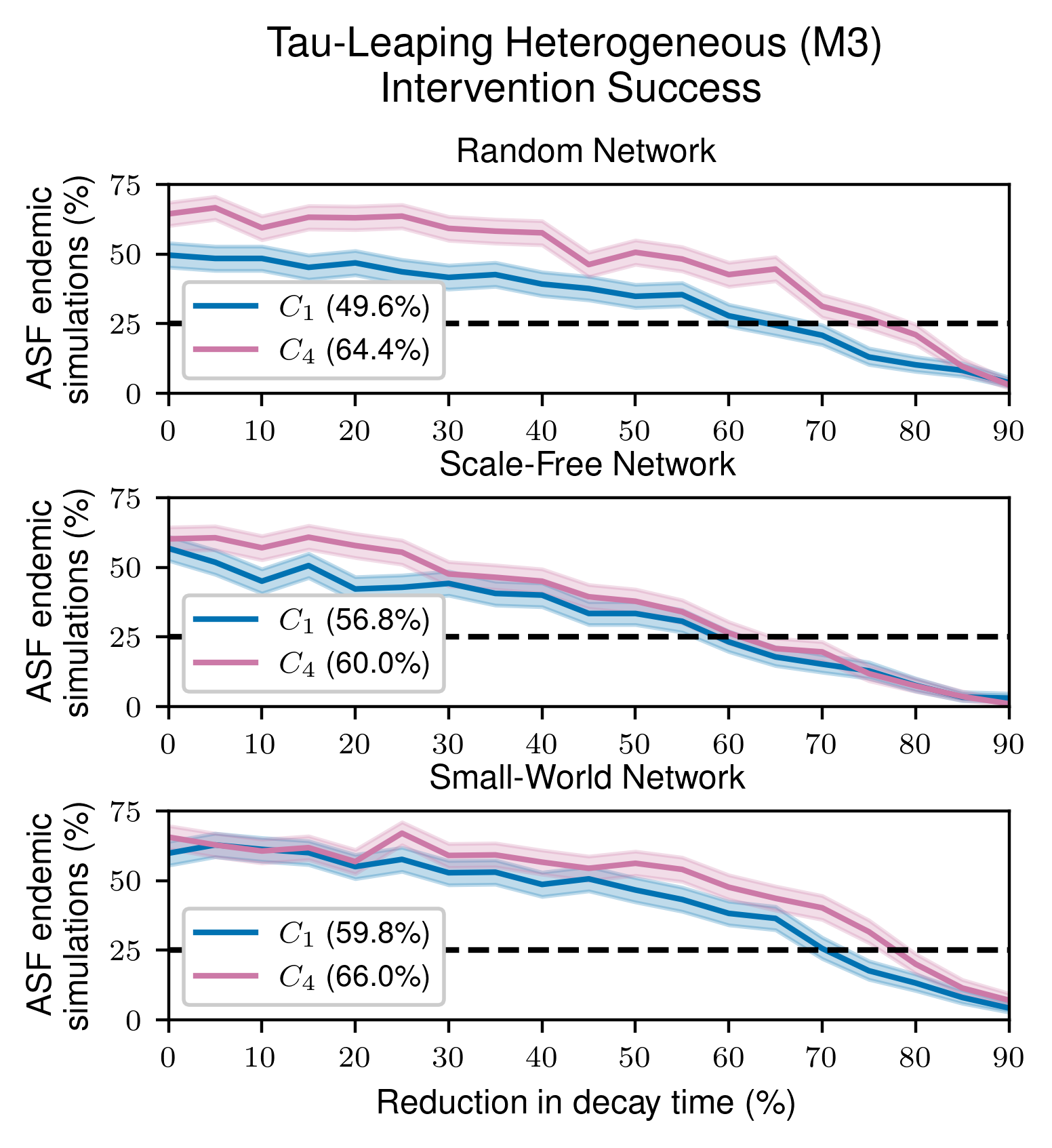}
	\caption{The effectiveness of a carcass removal intervention in the tau-leaping heterogeneous model (M3). Each row is the model run on a different network structure. In each plot the model investigates the impact of different assumed contact-density functions-- frequency-based ($C_1$) in blue and power-law based  ($C_4$) in blue. In brackets in the legend is the proportion of simulation where ASF is endemic with no intervention.}
	\label{fig:m3_inter}
\end{figure}

In M3, both network and density contact function type were found to influence intervention success. For the M3 interventions, the carcass decay period was reduced by a multiplicative factor $\alpha \in \left[0,0.05,\dots 0.85,0.9\right]$ and each magnitude of $\alpha$ was simulated 500 times. As with the previous analysis of M3, only $C_1$ and $C_4$ sub-models were included. All interventions were simulated for three years and began 90 days after ASFV was seeded in the population. For each network, after normalising for intervention-free endemicity, $C_1$ sub-models required comparable or lesser intensity of intervention to achieve elimination than required by the $C_4$ sub-models (Fig. \ref{fig:m3_inter}). The increased effectiveness of the intervention in $C_1$ M3 models, when compared to the $C_1$ homogeneous models, occurs as the $C_1$ M3 models have a greater relative proportion of transmission from carcasses.

Network choice influenced the likelihood of intervention success. We found that the SF network, on average, required the lowest intensity of intervention to reach the 75\% elimination benchmark, while SW and RN models required similar intensities of intervention. The better performance of the intervention on the SF network is unsurprising, due to the aforementioned SF network structure and the key role boar play in transmission. If a highly connected boar becomes infected, dies, and their carcasses are promptly removed via the intervention, this can quickly remove a highly connected node. This can disconnect regions of the network, limit spread and cause fade-out. Similar behaviour has been noted in previous scale-free studies~\parencite{keeling2005networks, Wang2003}. 

\section{Conclusion}
Previous wild boar focused ASF modelling studies have either been individual-based (agent-based) models~\parencite{gervasi2021african, pepin2020ecological, halasa2019simulation, lange2017elucidating, ko2023simulating} or, in the case of O`Niell \textit{et al.}~\parencite{o2020modelling}, an ODE model. This study investigates the gap between the two model types. To include a spatial component of transmission, while reducing complexity compared to an individual-based model, we used a stochastic ASF model on a network (M3). The network based approach has the ability to easily compare the effect of different network topologies on disease dynamics. This ability to analyse different networks is important due to the limited understanding of the social dynamics of wild pigs~\parencite{beasley2018research}. Similar to Lange \& Thulke~\parencite{lange2017elucidating}, Pepin \textit{et al.}~\parencite{pepin2020ecological}, and Gervasi \& Guberti~\parencite{gervasi2021african}, in our best fitting contact-density function, $C_4$, across all tested models and networks, we found that carcasses were essential to endemic transmission. Furthermore, similar to Pepin \textit{et al.}~\parencite{pepin2020ecological}, and Gervasi \& Guberti~\parencite{gervasi2021african}, we found that carcass removal was a viable intervention to control an ASF outbreak.

There are some limitations to the modelling conducted in this study that stem from the ongoing tug-of-war between complexity and simplicity. Firstly, to streamline all models, we did not model piglets, as this assumption allowed us to halve the number of classes. The addition of piglets would further increase the recruitment of susceptibles in the model during birthing periods, which could influence ASF dynamics and allow for more spread around these birthing periods. Similarly, we modelled yearlings and adult pigs as the same class. Yearlings and adult pigs have similar, but different characteristics and behaviours, for example, different litter sizes and mortality rates~\parencite{pascual2022wild}. Although carcass decay and births had a strong time dependence, the transmission rate ($\beta$), did not. Wild boar contact rates vary by season~\parencite{caley1993ecology}; therefore there should be a degree of seasonality to $\beta$. Time dependence in transmission has been shown to alter potential outbreak sizes, induce complex dynamics in endemic scenarios, and reduce disease persistence~\parencite{grassly2006seasonal}. 

There is a large degree of uncertainty surrounding a number of key processes in both the wild boar population and ASF spread dynamics. There is a knowledge gap on the group-level behaviour of wild boar; for example, the degree distribution of each group and the overall network characteristics. We attempted to allow for this by modelling ASF on three different networks; however, there are drawbacks with each chosen network. The SF networks have an untruncated degree distribution, which can cause some boar to have very large, unrealistic, degrees and being too influential in ASF dynamics. Furthermore, true scale-free networks are uncommon~\parencite{broido2019scale}. RN networks lack sufficient clustering and hubs, and most food webs do not exhibit small-world topology~\parencite{dunne2002food}. All networks in this study are static in both nodes and connections, which is likely unrealistic over the six year simulation period; however, the extent that a wild boar network evolves with time is unknown. Another limitation of the heterogeneous stochastic model (M3), is that after a boar dies, they remain connected to all their previously connected groups until their carcass decays. Boars were chosen to be the highly connected nodes on the basis that they have the largest home range, meaning a boar would travel and contact many other boar and sow groups. Once a boar dies, this travel cannot occur, but in our model the boar remains connected to the same number of other groups. Therefore, the model may overestimate the force of infection from boar carcasses, especially within the SF network.

The highly adaptable nature of the ASF modelling framework presented in this study allows it to be extended to model potential ASF incursions in different regions. However, to better model ASF, further research should be conducted on the aforementioned knowledge gaps. Moreover, as evidenced by this study and the recent ASF literature, wild boar carcasses play a key role in endemic ASF transmission. Therefore, to accurately model ASF in different geographic locations, it is crucial we understand the period of viability of ASFV in carcasses in different climatic conditions, e.g. humidity, temperature. The modelling framework could also be extended to include domestic pigs. As the interface between wild and domestic pigs is important factor of ASF outbreaks~\parencite{hayes2021mechanistic} and piggery nodes could include edges that allow for long distance transmission.

We found that potential ASF outbreak outcomes are highly dependent on model specification. Model complexity, density-contact function, network, and carrying capacity implementation all affect disease spread, persistence, and intervention effectiveness. Across all models, of all contact-density functions tested, we found that the power-law formulation ($C \propto \sqrt{\rho}$), was the most able to reproduce observed ASF outbreak behaviour. Models with this contact-density function involved substantial carcass-based transmission; however, the force of infection from carcasses varied between models, with increased model heterogeneity leading to a decrease in the relative importance of carcass-based transmission. We found these differences influenced the efficacy of potential interventions, which highlights the necessity of carefully considering model structure and the utility of using an ensemble of models.

\printbibliography

\appendix

\title{Modelling African Swine Fever in Wild Boar: Investigating Model Assumptions and Structure}

\setcounter{section}{0}
\setcounter{figure}{0}

	\section{Appendix}
	
	\subsection{Fitting the split equations}
	
	\begin{figure}[htp]
		\centering
		\includegraphics[]{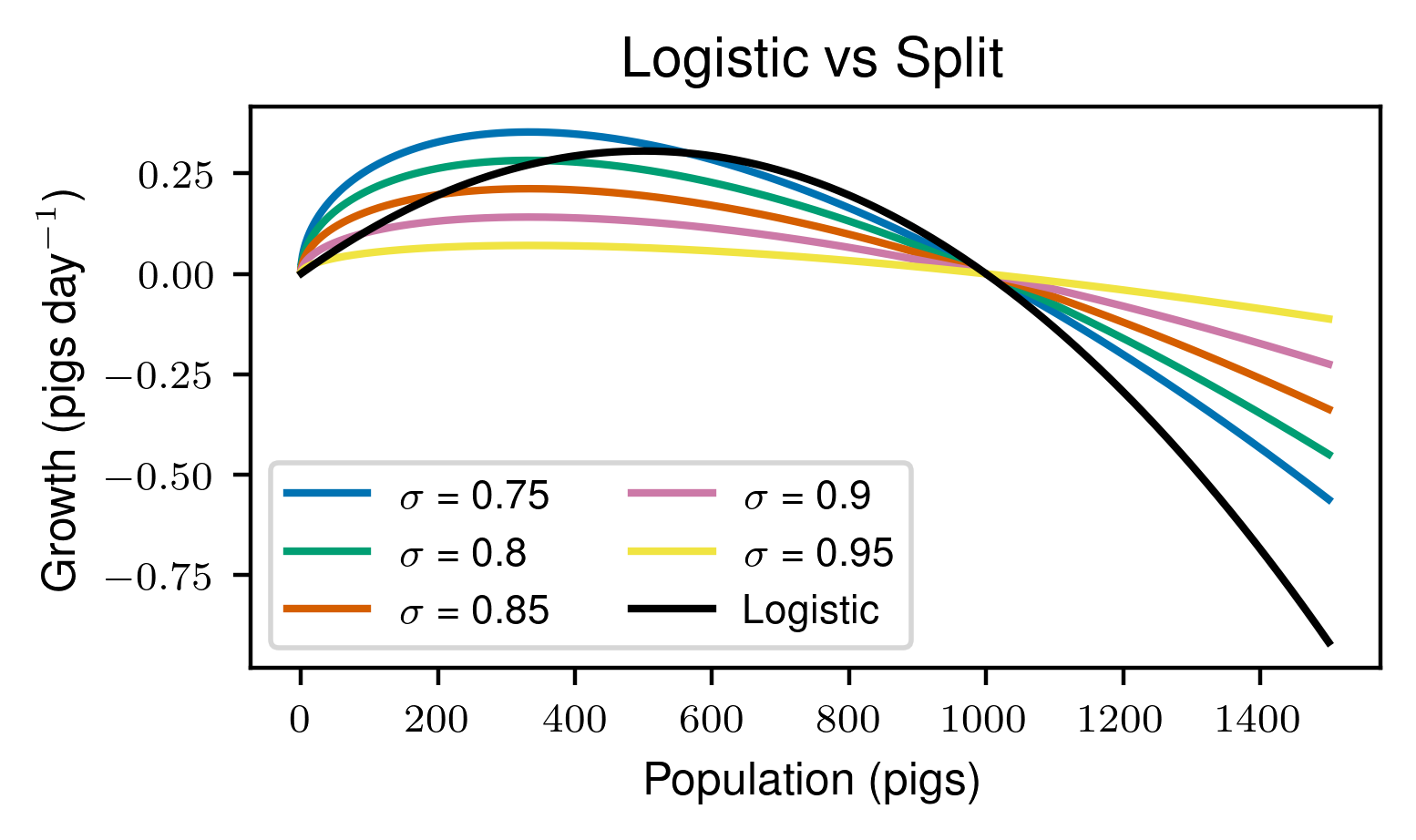}
		\caption{The effect of $\sigma$ on the net population growth rate for populations around the carrying capacity, where $\sigma$ is the ratio of density-independent births/deaths to total births/deaths (1 is purely density independent, whereas 0 is purely density-dependent). The carrying capacity of the population is 1000 pigs. The logistic model is included, shown by the black line, where the logistic model assumes that the death rate is two thirds of the birth rate.}
		\label{fig:sigmaeffects}
	\end{figure}
	
	Once the population size is below the carrying capacity ($N<K$), for the logistic equation to return towards $K$ in a period of time, the net growth rate over said period needs to be greater than one. Previous studies have estimated that the growth rate of wild boar in Europe, while uncertain, is between 0.85 and 1.63 \parencite{massei2015wild,pascual2022wild}. Therefore, we assumed a growth rate of 1.5. To implement this with the logistic equation we set yearly deaths to be two thirds of births. During an ASF outbreak, the population will primarily be between 0.25<$N/K$<1.0, therefore this is the domain we want the logistic equation and the split equations to behave similarly within. We found the split equation with $\sigma = 0.75$ offered the most similar behaviour to the logistic equation  (Fig. \ref{fig:sigmaeffects}) in this region. It is important to note that all realistic values of $\sigma$ for the split method offer a larger growth rate than the logistic equation when $N << K$. 
	
	\begin{figure}[h]
		\centering
		\includegraphics[]{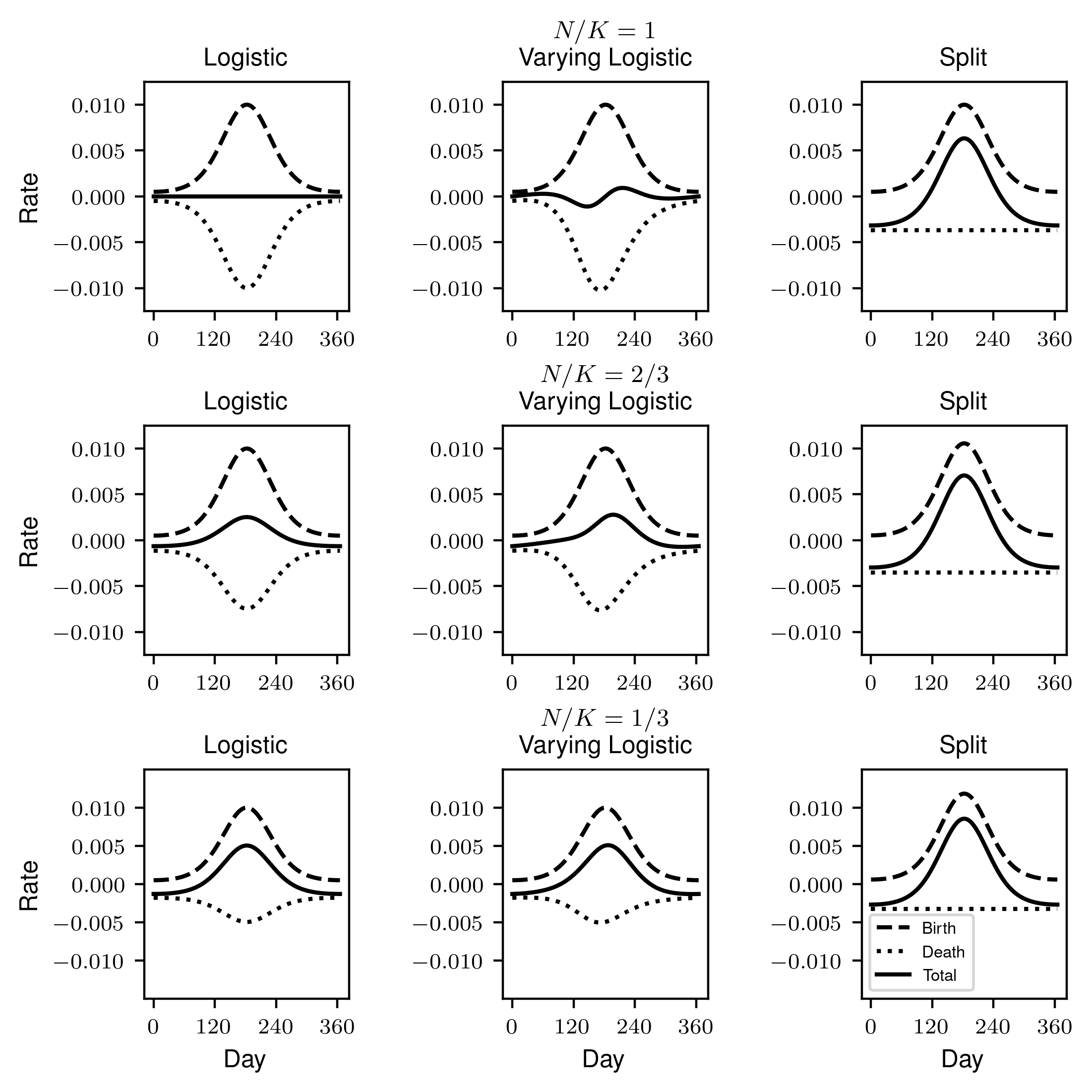}
		\caption{Comparison of the logistic model, the logistic model with seasonally varying $K$, and the \textit{split} method where births and deaths are decoupled ($\theta  =1/2, \sigma = 0.75$). The figure shows the net birth and death rates for each of the three models over a year, in the presence of a strong mid-year birth pulse ($k = 0.0136, s = 10, \phi_b = 0$). The rates were examined for three different ratios of $N/K$, where $N$ is the population size and $K$ is the carrying capacity.}
		\label{fig:timingissues}
	\end{figure}
	
	\clearpage
	\subsection{Logistic function}
	\label{logistic_section}
	\subsubsection{Equations}
	To gauge the effect of using the logistic function to implement the carrying capacity, instead of the split method outlined in the paper, we conducted the same fitting procedure for logistic equation versions of models M1 and M2 (now models LM1 and LM2). The addition of the logistic function changes Equation 2.6 to:
	\begin{align}
		\begin{split}
			\dfrac{dS}{dt}&= b(t) N + \kappa R - (\mu + g(t)N)S -\beta(I+\omega C)\frac{S}{N+C},\\
			\dfrac{dE}{dt}&=\beta(I+\omega C)\frac{S}{N+C}-(\mu +g(t)N+\zeta)E,\\
			\dfrac{dI}{dt}&=\zeta E - (\mu +g(t)N+\gamma)I,\\
			\dfrac{dR}{dt}&=\gamma(1-\nu)I - (\mu +g(t)N+\kappa)R,\\
			\dfrac{dC}{dt}&=(\mu +g(t)N + \gamma \nu)I -\lambda(t) C,\\
		\end{split}
		\label{eq:ode_sim_logistic}
	\end{align}
	where $b(t)$ is Equation 2.1 and $g(t)$ is equal to $(b(t)-\mu)/K$. All parameters are the same as used in the regular M1 and M2 models, apart from $\mu$, which equaled:
	\begin{equation}
		\mu = \frac{2}{3}\int_0^{365} \frac{b(t)}{365}dt.
	\end{equation}
	
	\subsubsection{Fitting}
\begin{figure}[h!]
		\centering
		\includegraphics{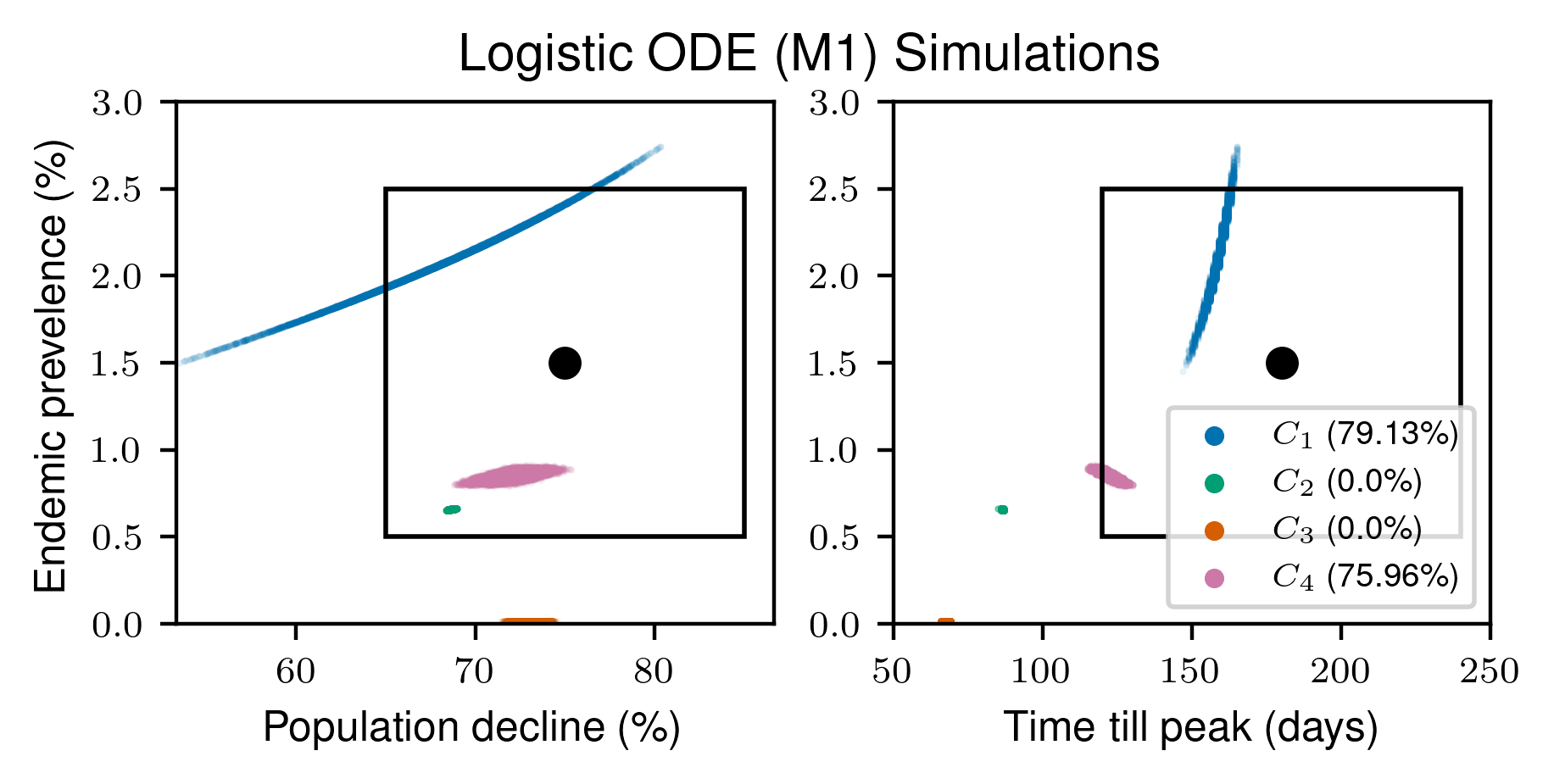}
		\caption{Comparison between the three simulated summary statistics and the observed summary statistics for the logistic-based ODE model (LM1). The plot on the left compares S1 and S2, while the right plot compares S1 and S3. The observed summary statistic is represented by the black dot, and its associated 95\% confidence interval by the black rectangle. Results from the frequency-based contact function ($C_1$) are in blue, from the density-based contact function ($C_2$) in green, from the sigmoid contact function ($C_3$) in orange, and from the power-law contact function ($C_4$) in pink. Each model was simulated 10,000 times. In brackets are the proportion of simulations where all three simulated summary statistics lie within the 95\% confidence interval of all observed summary statistics.}
		\label{fig:logodesims}
	\end{figure}

	Each contact-density function was simulated 10,000 times from the fitted posteriors, with the results of the fitting procedure for LM1 given in Fig. \ref{fig:logodesims}. Similar to M1, both the frequency-based ($C_1$) and the power-law based ($C_4$) density contact functions were able to reproduce the observed summary statistics, while the density-based ($C_2$) and sigmoid ($C_3$) density contact functions were unable to reproduce the summary statistics.
	
\begin{figure}[h!]
		\centering
		\includegraphics{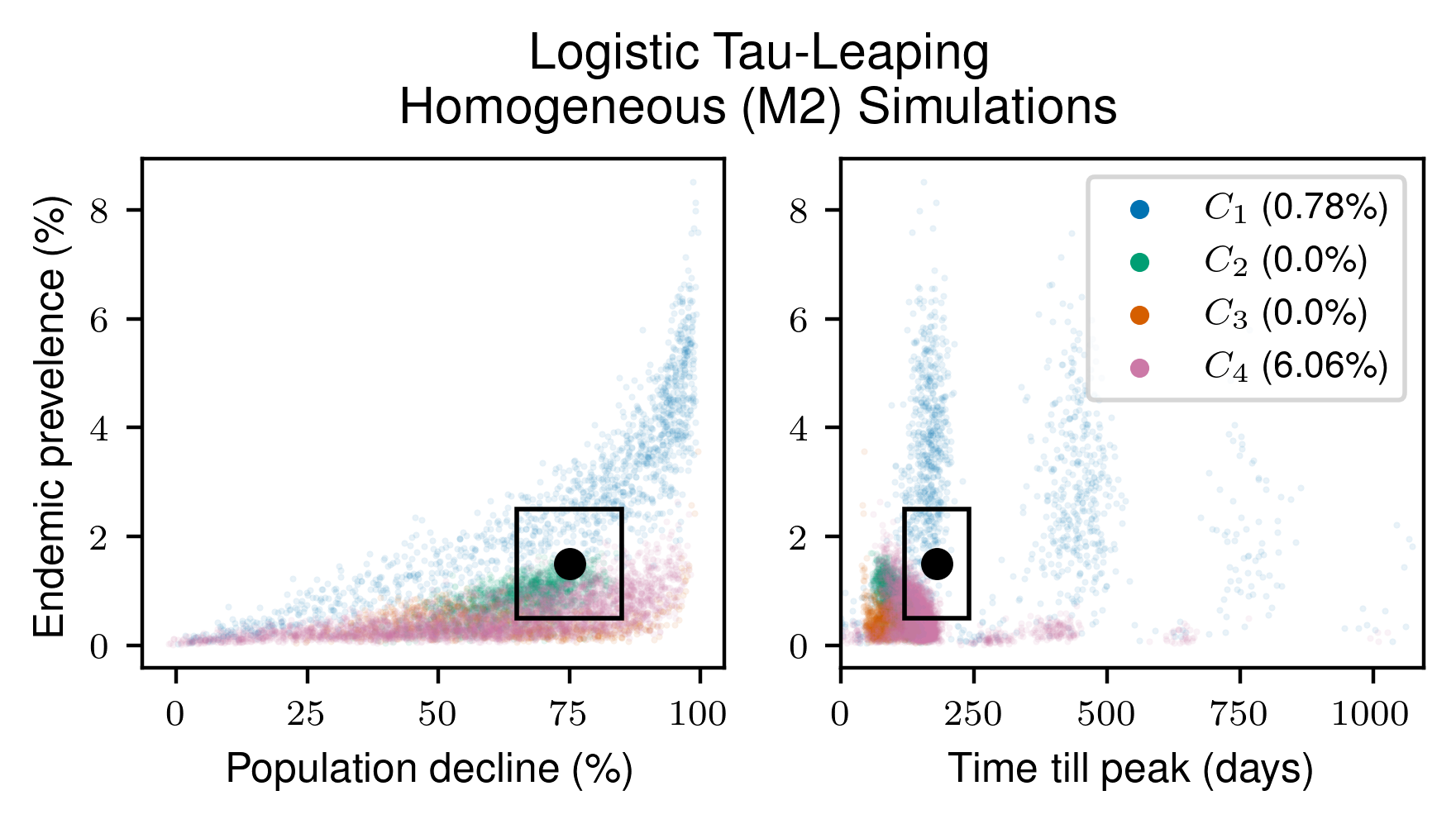}
		\caption{Comparison between the three simulated summary statistics and the observed summary statics for the logistic-based stochastic homogeneous model (LM2). The plot on the left compares S1 and S2, while the right plot compares S1 and S3. The observed summary statistic is represented by the black dot, and its associated 95\% confidence interval by the black rectangle. Results from the frequency-based contact function ($C_1$) are in blue, from the density-based contact function ($C_2$) in green, from the sigmoid contact function ($C_3$) in orange, and from the power-law contact function ($C_4$) in pink. Each model was simulated 10,000 times. In brackets are the proportion of simulations where all three simulated summary statistics lie within the 95\% confidence interval of all observed summary statistics.}
		\label{fig:logtauhsims}
	\end{figure}

	The aforementioned procedure was carried out with LM2, and the results are given in Fig. \ref{fig:logtauhsims}. Unlike M2, all transmission functions produced a poor quality of fit and were largely unable to replicate the observed summary statistics. Of the contact functions, $C_4$ was best able to reproduce the observed summary statistics, but only 6\% of simulations were within all 95\% confidence intervals.
	
	\subsubsection{Intervention}
\begin{figure}[h!]
		\centering
		\includegraphics{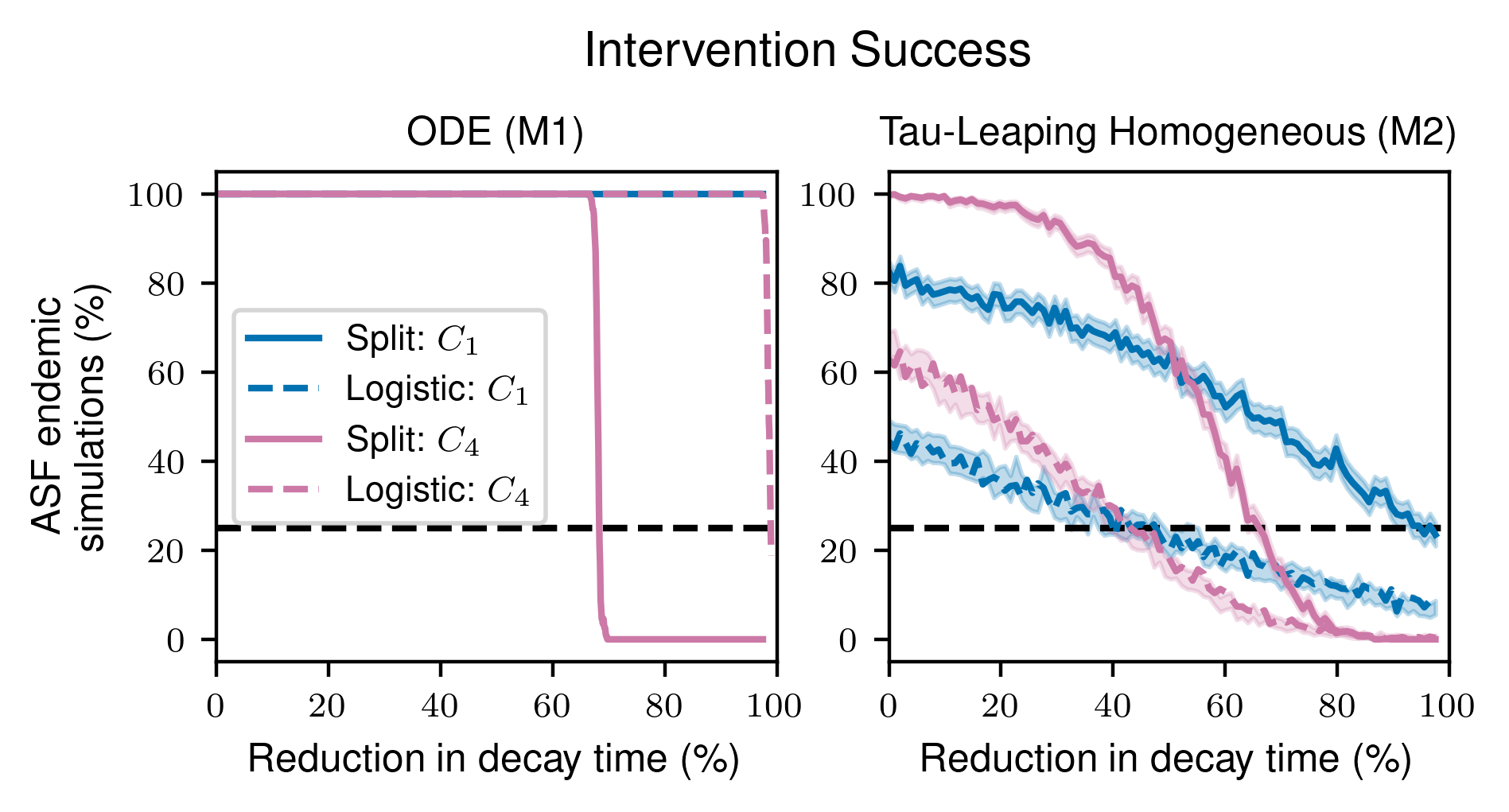}
		\caption{The effect of using the carrying capacity with the \textit{split} equation or logistic equation on intervention success. On the left if the ODE model and on the right is the tau-leaping homogeneous model. Two different density-contact functions were included, frequency-based ($C_1$) in blue and a power-law relationship ($C_4$) in pink. Solid lines are the simulations with the \textit{split} equations and dashed lines are simulations with the logistic equation. All interventions began 90 days after seeding ASFV in 1\% of the population and enedmicity was measured three years after the intervention began.}
		\label{fig:lm1lm2_inter}
	\end{figure}

	To assess the influence of using the logistic equation on ASF dynamics, we also simulated LM1 and LM2 with same intervention strategy used in M1 and M2. The carcass decay period ($1/\lambda$) was decreased by a factor ($\alpha$), where $\alpha \in [0,0.99]$. The intervention was only conducted with density-contact functions $C_1$ and $C_4$. The results of the intervention are given in Fig. \ref{fig:lm1lm2_inter}. Similar to M1, LM1 with $C_1$ predicted that carcass removal was an ineffective strategy as no magnitude of $\alpha$ could reduce endemicity. LM1 with $C_4$, had a bifurcation where the system switches from endemic solution to a non-endemic solution. This bifurcation required a significantly stronger intervention than what was required in M1. The transition occurred in M1 at $\alpha \approx 0.68$ and in LM1 at $\alpha \approx 0.96$. Intervention efficacy varied between M2 and LM2. As seen in Fig. \ref{fig:lm1lm2_inter}, LM2 endemicity when $\alpha = 0 $, for both $C_1$ and $C_4$ was significantly lower than was observed in M1. This lower endemicity stemmed from the poor fit of LM2. Over most values of $\alpha$, the slope of the LM2 simulations was shallower than that of the M2 simulations, highlighting the decreased effectiveness of the interventions in LM2.
	
	The differences in both the ability of the logistic models to replicate the observed summary statistics and the effectiveness of the intervention compared to M1 and M2 (which used the \textit{split} equation), highlight the importance of considering how carrying capacity is implemented in simple models. 
	
	\clearpage
	
	\subsection{Tau-leaping Processes}
	
	\begin{table}[h!]
		
		\centering
		\begin{tabular}{ m{6cm}   m{5cm}  m{2.5cm}} \toprule
			
			Rate & Description & Effect  \\\midrule
			
			$f_b(t, \mathbf{N})N_{j}$ & Birth  & $S_{j}\rightarrow S_{j} + 1$ \\[0.35cm]
			
			$\sum^{n}_{\substack{k=1 \\ k \neq j}}\beta_{jk}(I_k+\omega C_k)\frac{S_j}{N_j+ C_j +N_k +C_k}$\newline $-\beta_{jj}(I_j+\omega C_j)\frac{S_j}{N_j+C_j}$ & Infection & $S_{j}\rightarrow S_{j} - 1$ \newline $E_{j}\rightarrow E_{j} + 1$ \\[0.35cm] 
			
			$\zeta E_j$ & Infectious & $E_j\rightarrow E_j - 1$ \newline $I_j\rightarrow I_{j} +1$ \\[0.35cm]
			
			$\gamma(1-\nu)I_{j}$ & Recovery & $I_{j}\rightarrow I_{j} - 1$ \newline $R_{j}\rightarrow R_{j} + 1$ \\[0.35cm]
			
			$\gamma\nu I_{j}$ & ASF related death & $I_{j}\rightarrow I_{j} - 1$ \newline $C_{j}\rightarrow C_{j} + 1$ \\[0.35cm] 
			
			$\kappa R_{j}$ & Waning of immunity & $R_{j}\rightarrow R_{j} - 1$ \newline $S_{j}\rightarrow S_{j} + 1$ \\[0.35cm]
			
			$\lambda(t) C_{j}$ & Carcass decay & $C_{j}\rightarrow C_{j} - 1$ \\[0.35cm]
			
			$f_d(\mathbf{N}) S_j$ & Natural death in S & $S_{j}\rightarrow S_{j} - 1$ \\[0.35cm]
			
			$f_d(\mathbf{N}) E_j$ & Natural death in E & $E_{j}\rightarrow E_{j} - 1$ \\[0.35cm]
			
			$f_d(\mathbf{N}) I_j$ & Natural death in I &  $I_{j}\rightarrow I_{j} - 1$ \newline $C_{j}\rightarrow C_{j} + 1$  \\[0.35cm]
			
			$f_d(\mathbf{N}) R_j$ & Natural death in R &  $R_{j}\rightarrow R_{j} - 1$ \\[0.35cm]\bottomrule
			
		\end{tabular}
		\caption{The eleven discrete tau-leaping model processes for group $j$ in a single population. These processes are used in both the network and homogeneous models, however, in the homogeneous model (M2) there is only a single group that represents the entire population.}
		\label{table:tau_reactions}
	\end{table} 
	\clearpage

	\subsection{Posterior plots}
	\begin{figure}[ht!]
		\centering
		\includegraphics{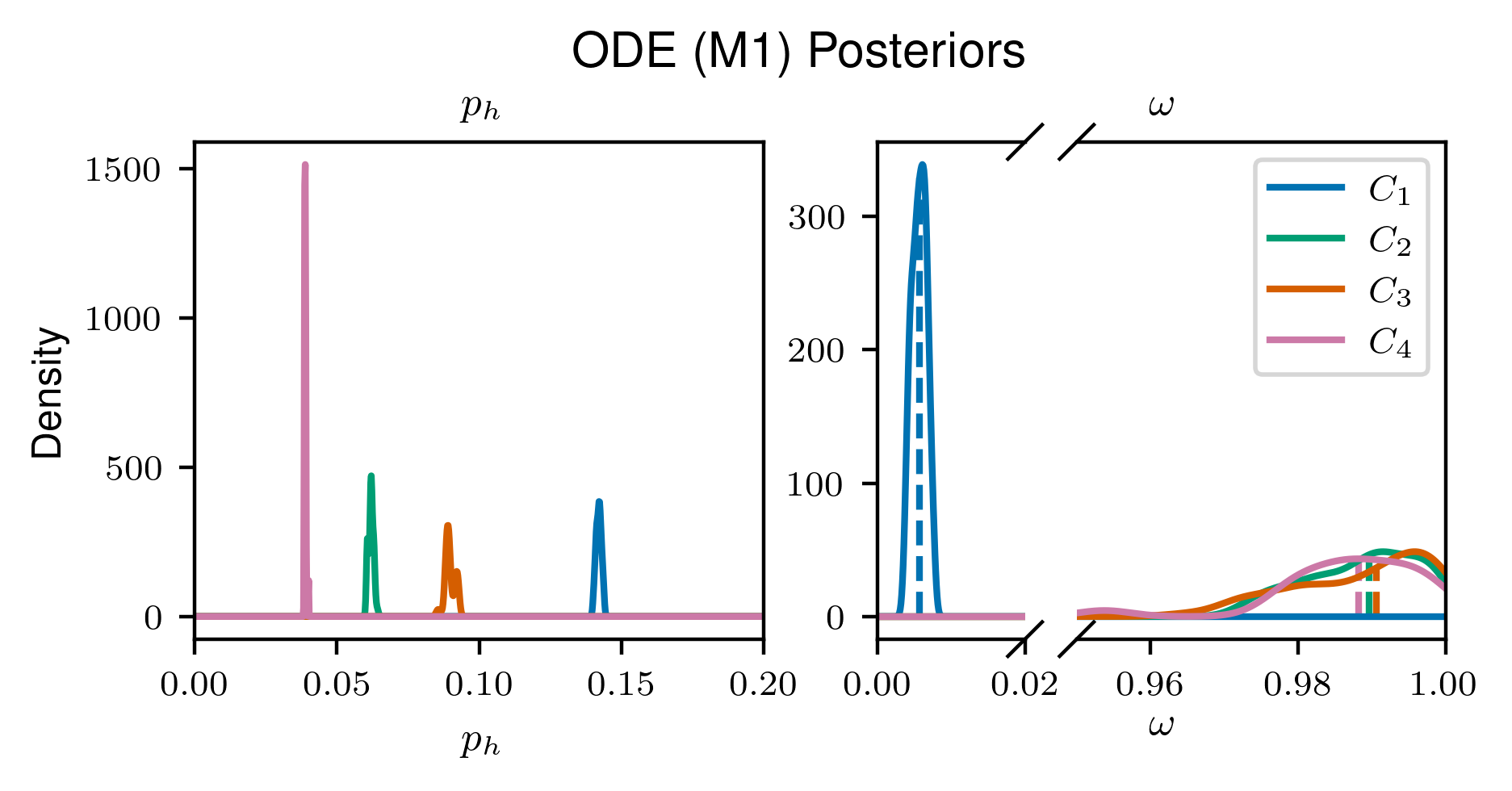}
		\caption{The final generation kernel-smoothed posterior densities from the ABC-SMC fitting of the ODE model (M1) for each of the four contact-density functions. The left plot shows the posteriors of the contact dependent transmission-coefficient ($p_h$) and the right plot shows the posteriors of the carcass transmission modifier ($\omega$). In blue is the model with frequency-based transmission ($C_1$), in green is the model with pure density-based transmission ($C_2$), in orange is the model with sigmoid relationship ($C_3$), and power-law in pink ($C_4$).}
		\label{fig:m1postplot}
	\end{figure}
	
	\begin{figure}[h!]
		\centering
		\includegraphics{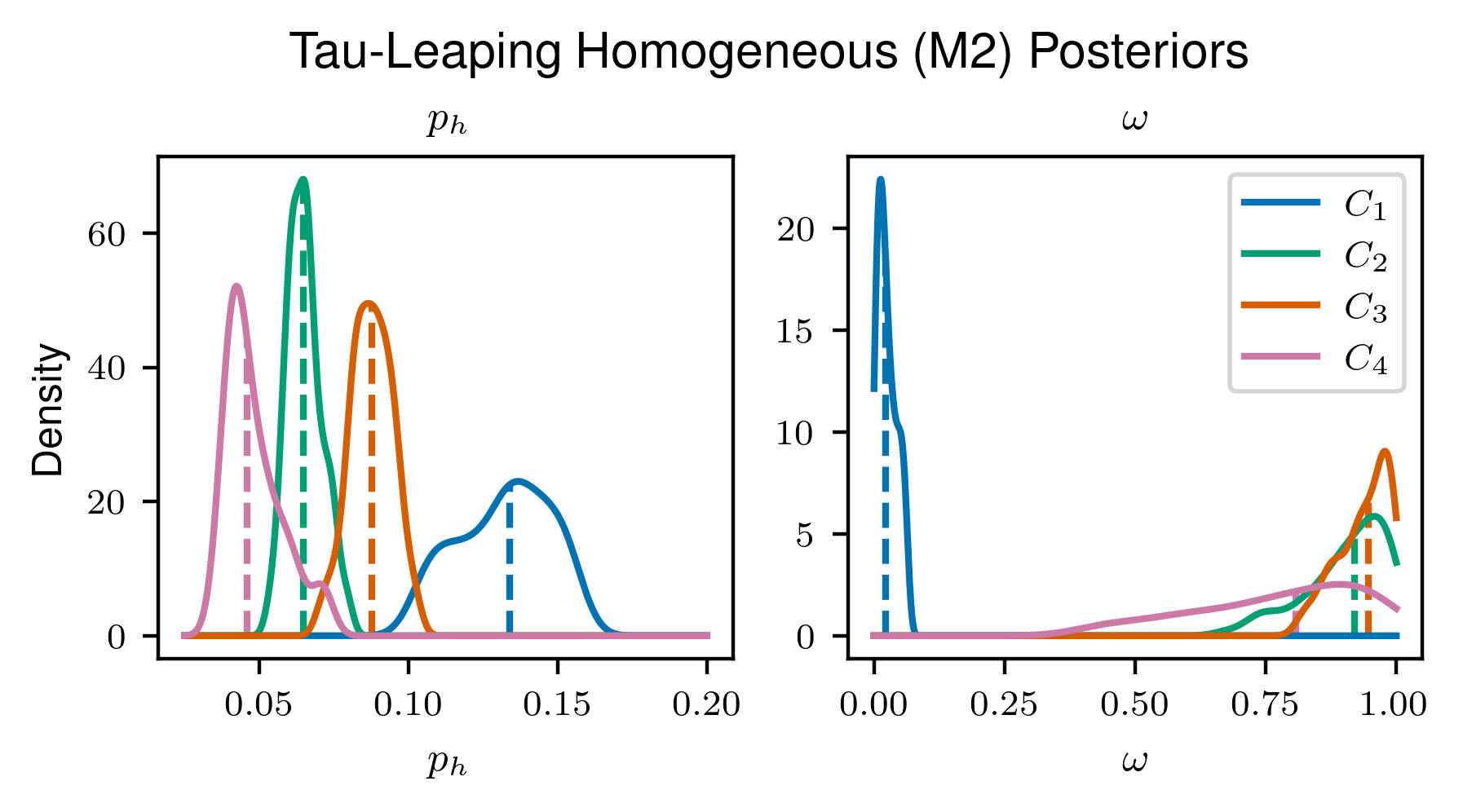}
		\caption{The final generation kernel-smoothed posterior densities from the ABC-SMC fitting of the homogeneous tau-leaping model (M2) for each of the four contact-density functions. The left plot shows the posteriors of the contact dependent transmission-coefficient ($p_h$) and the right plot shows the posteriors of the carcass transmission modified ($\omega$). In blue is the model with frequency-based transmission ($C_1$), in green is the model with pure density-based transmission ($C_2$), in orange is the model with sigmoid relationship ($C_3$), and power-law in pink ($C_4$).}
		\label{fig:m2postplot}
	\end{figure}
	
	\begin{table}[h!]
		\centering
		\begin{tabular}{llcccccc}\toprule
			& & \multicolumn{3}{c}{$p_h$} & \multicolumn{3}{c}{$\omega$}\\ \cmidrule(lr){3-5}\cmidrule(lr){6-8}
			Model & Contact Type & MAP & Mean & Median & MAP & Mean & Median\\\midrule
			
			M1 & $C(\rho_N) =  1$ & 0.142 &0.142 &0.142 &0.006 &0.006 &0.006 \\
			& $C(\rho_N) \propto  \rho_N$ & 0.062 &0.062 &0.062 &0.991 &0.989 &0.990    \\ 
			& $C(\rho_N) \propto  \tanh(\rho_N)$ & 0.089 &0.090 &0.089 &0.996 &0.989 &0.991   \\ 
			& $C(\rho_N) \propto  \sqrt{\rho_N }$ & 0.039 &0.039 &0.039 &0.988 &0.987 &0.988  \\ \midrule
			
			M2& $C(\rho_N) =  1$ & 0.137 &0.132 &0.133 &0.013 &0.025 &0.022  \\ 
			& $C(\rho_N) \propto   \rho_N$ & 0.065 &0.065 &0.065 &0.958 &0.904 &0.921  \\ 
			& $C(\rho_N) \propto  \tanh(\rho_N)$ & 0.087 &0.088 &0.088 &0.978 &0.937 &0.947  \\ 
			&$C(\rho_N) \propto  \sqrt{\rho_N}$ & 0.042 &0.048 &0.046 &0.892 &0.778 &0.805   \\  \bottomrule
		\end{tabular}
		\caption{The mean, median, and maximum a posteriori probability (MAP) of fitted parameters from the ODE model (M1) and tau-leaping homogeneous model (M2). Here, $p_h$ is the homogeneous transmission rate assuming contact and $\omega$ is the relative infectiousness  of carcasses to live pigs. }
		\label{table:propm1m2}
	\end{table} 
	
	\begin{figure}[b]
		\centering
		\includegraphics{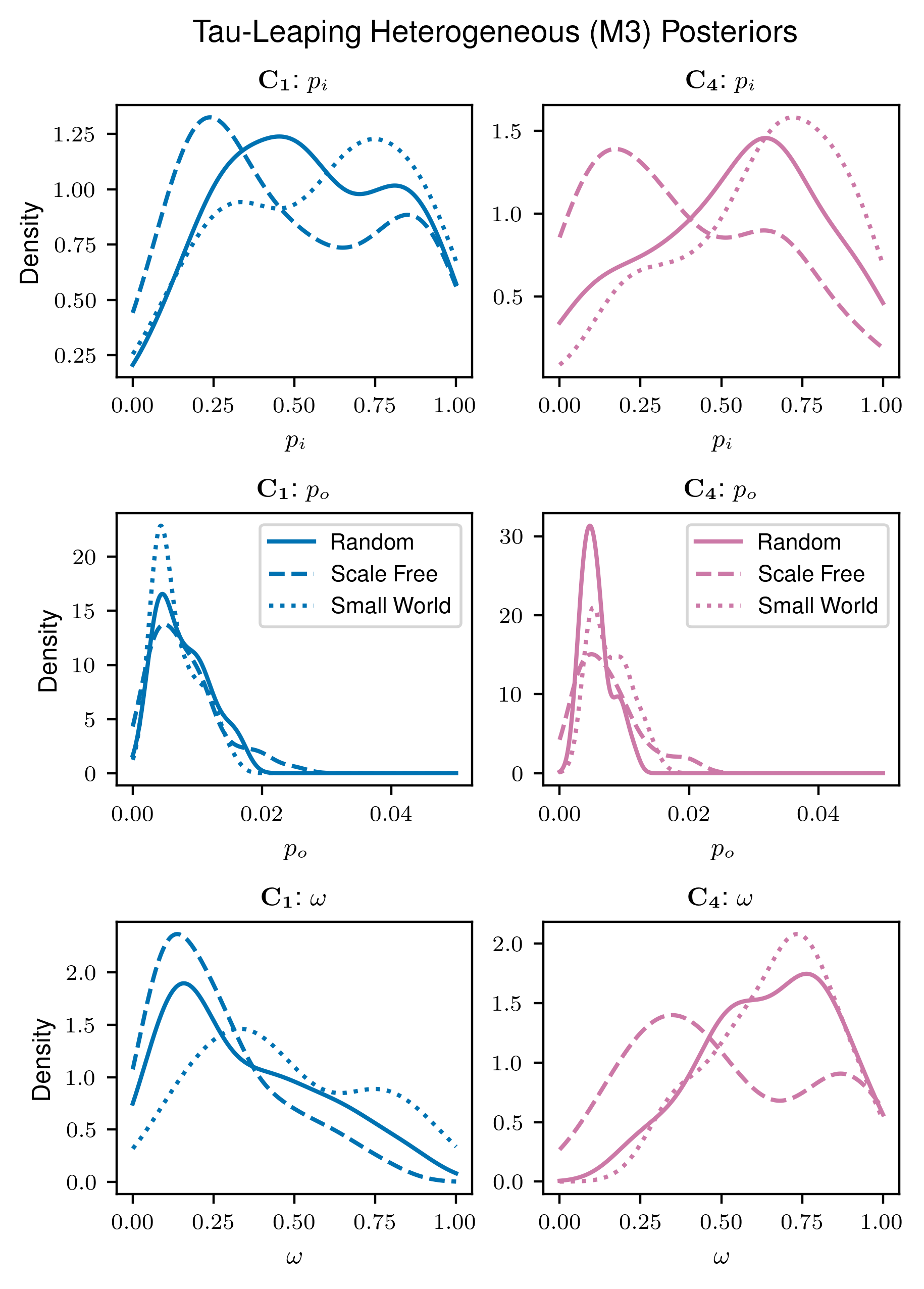}
		\caption{The final generation kernel-smoothed posterior densities from the ABC-SMC fitting of the heterogeneous tau-leaping network model (M3) for the two most accurate contact-density functions: frequency ($C_1$) and power-law ($C_4$). Each row is a different fitted parameter, the top row is the contact dependent intra-group transmission coefficient $(p_i)$, the middle row is the contact dependent inter-group transmission coefficient $(p_o)$, and the bottom row is the carcass transmission modified ($\omega$). The plots on the left, in blue, are from the $C_1$ sub-model, while the plots on the right, in pink, are from the $C_4$ sub-model. Each plot contains the posteriors from each of the three tested network types, the solid line is a random (RN) network, the large-dash line is a scale-free (SF) network, and the small-dash line is the small-world (SW) network.}
		\label{fig:m3postplot}
	\end{figure}
	
	\begin{table}[h!]
		\centering
		\begin{tabular}{llccccccccc}\toprule
			& & \multicolumn{3}{c}{$p_i$} & \multicolumn{3}{c}{$p_o$} & \multicolumn{3}{c}{$\omega$}\\ \cmidrule(lr){3-5}\cmidrule(lr){6-8}\cmidrule(lr){9-11}
			Network & Contact Type & MAP & Mean & Median & MAP & Mean & Median & MAP & Mean & Median \\\midrule
			
			RN & $C(\rho_N) =  1$ & 0.453 &0.548 &0.536 & 0.0046 &0.0078 & 0.0071 &0.158 &0.345 &0.295  \\ 
			& $C(\rho_N) \propto  \sqrt{\rho_N }$ & 0.637 &0.545 &0.572 & 0.0047 & 0.0057 & 0.0053 & 0.764 &0.652 &0.667  \\ \midrule
			
			SF & $C(\rho_N) =  1$ & 0.239 &0.484 &0.439 & 0.0049 &0.0079 & 0.0070 &0.138 &0.261 &0.218 \\ 
			& $C(\rho_N) \propto  \sqrt{\rho_N }$ & 0.173 &0.376 &0.335 &0.0049 &0.0074 &0.0065   &0.348 &0.516 &0.468 \\ \midrule
			
			SW & $C(\rho_N) =  1$ & 0.747 &0.578 &0.603 &0.0044 &0.0067 &0.0058  &0.332&0.472 &0.435 \\ 
			& $C(\rho_N) \propto  \sqrt{\rho_N }$ & 0.721 &0.630 &0.663 & 0.0052 & 0.0079 & 0.0074 &0.734 &0.667 &0.686  \\ \bottomrule
		\end{tabular}
		\caption{The mean, median, and maximum a posteriori probability (MAP) of fitted parameters from the heterogeneous tau-leaping model (M3). Fitting was conducted on Erd{\H{o}}s–R{\'e}nyi random network (RN), Barab{\'a}si–Albert scale-free network (SF), and Watts–Strogatz small-world random network (SW).  Here, $p_i$ is the intra-group transmission rate assuming contact, $p_o$ is the inter-group transmission rate assuming contact  and $\omega$ is the relative infectiousness  of carcasses to live pigs.}
		\label{table:propm3}
	\end{table} 
	
	\clearpage
	\subsection{Force of infection}
	\begin{table}[h!]
		\centering
		\begin{tabular}{llll}\toprule
			
			Contact type & Force of infection &  M1& M2  \\\midrule
			
			$C(\rho_N) =  1$ & Live  & $0.6825 \; \left[0.6787-0.6862\right]$ & $ 1.285 \; \left[1.266-1.303\right]$  \\
			& Carcass & $0.0307 \; \left[0.0302-0.0312\right]$ & $ 0.427 \; \left[0.416-0.438\right]$ \\
			& Total  & $0.7131 \; \left[0.7089-0.7173\right]$  & $ 1.712 \; \left[1.683-1.740\right]$ \\
			
			$C(\rho_N) \propto  \sqrt{\rho_N }$ & Live  & $0.0700 \; \left[0.0698-0.0701 \right]$ & $ 0.084 \; \left[0.083-0.085\right]$\\
			& Carcass & $ 0.4414 \; \left[0.4403-0.4424\right]$ & $ 0.456 \; \left[0.450-0.462\right]$\\
			& Total  & $0.5113 \; \left[0.5101-0.5125\right]$  & $ 0.540 \; \left[0.533-0.547\right]$ \\\bottomrule
		\end{tabular}
		
		\caption{The yearly force of infection from carcasses and live boar. From the ODE model (M1) and the homogeneous tau-leaping model (M2), with both frequency-dependent transmission ($C_1$) and power-law contact-density function ($C_4$). Results were filtered to only include endemic simulations. }
		\label{table:forcem1m2}
	\end{table}

	\begin{table}[h!]
		\centering
		\begin{tabular}{lllll}\toprule
			Contact type & Source &  RN & SF & SW\\\midrule
			
			$C(\rho_N) =  1$ & Live boar & $ 0.044 \; \left[0.041-0.047\right]$ 
			& $ 0.216 \; \left[0.204-0.227\right]$ 
			& $ 0.022 \; \left[0.02-0.023\right]$  \\
			
			& Live Sows & $ 0.078 \; \left[0.074-0.083\right]$ 
			& $ 0.074 \; \left[0.07-0.078\right]$
			& $ 0.084 \; \left[0.079-0.089\right]$ \\
			& Carcass boar & $ 0.064 \; \left[0.060-0.068\right]$ 
			& $ 0.281 \; \left[0.265-0.298\right]$ 
			& $ 0.047 \; \left[0.044-0.050\right]$ \\
			& Carcass Sows & $ 0.121 \; \left[0.114-0.128\right]$
			& $ 0.117 \; \left[0.110-0.124\right]$
			& $ 0.195 \; \left[0.185-0.206\right]$ \\
			& Total & $ 0.307 \; \left[0.291-0.323\right]$
			& $ 0.688 \; \left[0.655-0.722\right]$
			& $ 0.348 \; \left[0.33-0.365\right]$ \\
			
			$C(\rho_N) \propto  \sqrt{\rho_N }$ & Live boar & $ 0.018 \; \left[0.017-0.019\right]$
			& $ 0.103 \; \left[0.098-0.109\right]$
			& $ 0.014 \; \left[0.013-0.015\right]$  \\
			& Live Sows & $ 0.030 \; \left[0.029-0.031\right]$
			& $ 0.033 \; \left[0.031-0.035\right]$
			& $ 0.052 \; \left[0.05-0.054\right]$ \\
			& Carcass boar & $ 0.059 \; \left[0.056-0.062\right]$
			& $ 0.237 \; \left[0.224-0.249\right]$
			& $ 0.044 \; \left[0.042-0.046\right]$\\
			& Carcass Sows & $ 0.101 \; \left[0.096-0.105\right]$
			& $ 0.087 \; \left[0.083-0.092\right]$ 
			& $ 0.175 \; \left[0.168-0.181\right]$ \\
			& Total & $ 0.208 \; \left[0.199-0.216\right]$
			& $ 0.460 \; \left[0.438-0.483\right]$
			& $ 0.285 \; \left[0.274-0.296\right]$ \\ \bottomrule
		\end{tabular}
		
		\caption{The force of infection from the tau-leaping heterogeneous model (M3). Force of infection calculated for live boar, live sows, boar carcasses. and sow carcasses. Modelled with both frequency-dependent transmission ($C_1$) and power-law contact-density function ($C_4$). Results were filtered to only include endemic simulations.}
		\label{table:forcem3}
	\end{table}  
	\clearpage
	\subsection{Basic reproduction number}
\begin{figure}[h]
		\centering
		\includegraphics{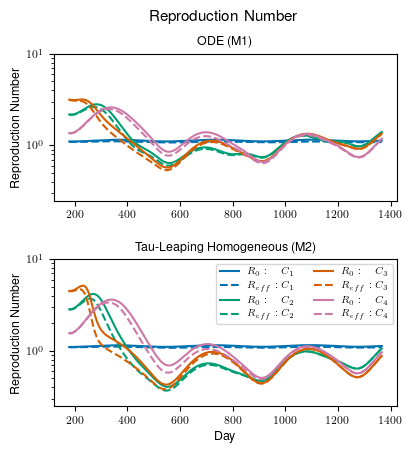}
		\caption{The change in both basic reproduction number ($R_0$) and effective reproduction number ($R_{eff}$) over a three year cycle for the Baltic region for each contact-density function type. The ODE model (M1) is shown in the top plot and the homogeneous tau-leaping model (M2) is shown in the bottom plot.}
		\label{fig:m1rep}
	\end{figure}

	Using the next generation matrix \parencite{diekmann2010construction} we were able to derive the basic reproduction number ($R_0$) for the ODE model (M1) and homogeneous tau-leaping model (M2):
	\begin{equation}
		R_0 = \frac{\beta_h \zeta (\lambda + \omega(f_d(L)+\nu\gamma))}{\lambda(f_d(L)+\gamma)(f_d(L)+\zeta)}
	\end{equation}
	We then investigated how $R_0$ and $R_{eff}$ varied seasonally over three years for each of the four contact-density functions, for an ASF outbreak in the Baltic region. The results for both model types are shown in Fig. \ref{fig:m1rep}.
	
	As seen in Fig. \ref{fig:m1rep}, the reproduction numbers for each contact-density function type are very similar between M1 and M2. The $C_2-C_4$ sub-models displayed a strong seasonal cycle in $R$, while $C_1$ sub-models do not exhibit this behaviour. This seasonal cycle is caused by infection from decaying carcasses as the decay rate varies seasonally. This effect is not as pronounced with the $C_1$ sub-models as carcass-based transmission was found to be negligible. Interestingly, this seasonal cycle found in $C_2 - C_4$ sub-models, in both M1 and M2, causes both $R_0$ and $R_{eff}$ to fall below 1 during the summer when the decay rate is highest and increase above 1 during the colder months when decay is slower.
	\clearpage

\end{document}